\documentclass[aps,twocolumn,a4paper,floatfix,superscriptaddress,pra,longbibliography]{revtex4-2}
\usepackage[toc,page]{appendix}
\usepackage{braket}
\usepackage{epsfig,graphicx,times}
\usepackage{amstext}
\usepackage{amsmath}
\usepackage{amssymb}
\usepackage{mathrsfs}
\usepackage{dcolumn}
\usepackage{bm}
\usepackage[tight]{subfigure}
\usepackage[colorlinks,linkcolor=blue,anchorcolor=blue,citecolor=blue,urlcolor=blue]{hyperref}
\usepackage{color}
\usepackage{siunitx}
\usepackage{appendix}
\usepackage{placeins}
\usepackage{textcomp}
\usepackage{bbold}
\usepackage{float}
\usepackage{verbatim}
\usepackage{minitoc}
\usepackage[dvipsnames]{xcolor}

\usepackage{soul}
\usepackage[framemethod=tikz]{mdframed}
\usepackage{lipsum}
\usepackage{epstopdf}

\begin{document}

\title{Tripartite quantum entanglement with squeezed optomechanics}

\author{Ya-Feng Jiao}
\affiliation{School of Electronics and Information, Zhengzhou University of Light Industry, Zhengzhou 450001, China}
\affiliation{Academy for Quantum Science and Technology, Zhengzhou University of Light Industry, Zhengzhou 450002, China}

\author{Yun-Lan Zuo}
\affiliation{Key Laboratory of Low-Dimensional Quantum Structures and Quantum Control of Ministry of Education, \\
Department of Physics and Synergetic Innovation Center for Quantum Effects and Applications, \\ Hunan Normal University, Changsha 410081, China}

\author{Yan Wang}
\affiliation{School of Electronics and Information, Zhengzhou University of Light Industry, Zhengzhou 450001, China}
\affiliation{Academy for Quantum Science and Technology, Zhengzhou University of Light Industry, Zhengzhou 450002, China}

\author{Wangjun Lu}
\affiliation{Institute of Engineering Education and Engineering Culture Innovation and Department of Maths and Physics, Hunan Institute of Engineering, Xiangtan 411104, China}

\author{Jie-Qiao Liao}
\affiliation{Key Laboratory of Low-Dimensional Quantum Structures and Quantum Control of Ministry of Education, \\
Department of Physics and Synergetic Innovation Center for Quantum Effects and Applications, \\ Hunan Normal University, Changsha 410081, China}

\author{Le-Man Kuang}\email{lmkuang@hunnu.edu.cn}
\affiliation{Academy for Quantum Science and Technology, Zhengzhou University of Light Industry, Zhengzhou 450002, China}
\affiliation{Key Laboratory of Low-Dimensional Quantum Structures and Quantum Control of Ministry of Education, \\
Department of Physics and Synergetic Innovation Center for Quantum Effects and Applications, \\ Hunan Normal University, Changsha 410081, China}

\author{Hui Jing}\email{jinghui73@gmail.com}
\affiliation{Academy for Quantum Science and Technology, Zhengzhou University of Light Industry, Zhengzhou 450002, China}
\affiliation{Key Laboratory of Low-Dimensional Quantum Structures and Quantum Control of Ministry of Education, \\
Department of Physics and Synergetic Innovation Center for Quantum Effects and Applications, \\ Hunan Normal University, Changsha 410081, China}

\date{\today}

\begin{abstract}
The ability to engineer entangled states that involve macroscopic objects is of particular importance for a wide variety of quantum-enabled technologies, ranging from quantum information processing to quantum sensing. Here we propose how to achieve coherent manipulation and enhancement of quantum entanglement in a hybrid optomechanical system, which consists of a Fabry-P\'{e}rot cavity with two movable mirrors, an optical parametric amplifier (OPA), and an injected squeezed vacuum reservoir. We show that the advantages of this system are twofold: (i) one can effectively regulate the light-mirror interactions by introducing a squeezed intracavity mode via the OPA; (ii) when properly matching the squeezing parameters between the squeezed cavity mode and the injected squeezed vacuum reservoir, the optical input noises can be suppressed completely. These peculiar features of this system allow us to generate and manipulate quantum entanglement in a coherent and controllable way. More importantly, we also find that such controllable entanglement, under some specific squeezing parameters, can be considerably enhanced in comparison with those of the conventional optomechanical system. Our work, providing a promising method to regulate and tailor the light-mirror interaction, are poised to serve as a useful tool for engineering various quantum effects which are based on cavity optomechanics.
\end{abstract}

\maketitle

\section{Introduction}

Entanglement, as a unique feature of quantum mechanics, is an appealing resource that enables a wide range of emerging quantum technologies~\cite{Horodecki2009RMP}. For example, entangled resources have been required to implement complex calculations on quantum computers~\cite{Zhang2018NSR}, to perform quantum communication tasks with security~\cite{XuRMP2020,Kimble2008Nature,Kuzyk2018PRX}, and to circumvent the standard quantum limit for quantum sensing~\cite{Degen2017RMP,Fleury2015NC,Zhao2020SCMPA}. So far, a great deal of efforts have been devoted to generating, manipulating and storing entanglement from theories to experiments~\cite{Friis2019NRP}, which involve photons~\cite{Yao2012NP,Wang2016PRL}, ions~\cite{Haffner2005Nature,Stute2012Nature}, atomic ensembles~\cite{Dai2016NP,Karg2020Science,Luo2022PRL}, and superconducting circuits~\cite{Campagne2018PRL,Kurpiers2018Nature}. Moreover, with the rapid advances on ground-state cooling of mechanical oscillators~\cite{Connell2010Nature,Lai2018PRA,Huang2022PRA}, the recent exploration of entanglement has also been extended to the domain of cavity optomechanics (COM)~\cite{Aspelmeyer2014RMP,Xiong2015SCPMA,Barzanjeh2022NP}, achieving macroscopic entangled states between light and motion~\cite{Vitali2007PRL,Palomaki2013Science,Ghobadi2014PRL,Riedinger2016Nature,Simon2018PRL}, between optical radiation fields~\cite{Tan2011PRA,Barzanjeh2019Nature,Chen2020NC}, and between mechanical vibrations~\cite{Hartmann2008PRL,Liao2014PRA,Li2015NJP,Korppi2018Nature,Riedinger2018Nature,Kotler2021Science,Mercier2021Science}. In very recent experiments, by exploiting the multi-tone driving field as an intermediary, direct observation of deterministic entanglement generation was even demonstrated between two massive mechanical oscillators with high fidelity~\cite{Kotler2021Science,Mercier2021Science}. Beyond these classes of bipartite entanglement, quantum entanglement shared by more than two partitions can play a more essential role in the development of programmable quantum networks\cite{Armstrong2012NC,Cai2017NC} and controllable dense coding\cite{Jing2003PRL}. Recently, by integrating another degree of freedom to COM system, several promising schemes have also been proposed theoretically for achieving full tripartite entanglement, which includes magnon-photon-phonon entanglement~\cite{Li2018PRL}, atom-light-mirror entanglement~\cite{Genes2008PRA}, and light-mirror-mirror entanglement~\cite{Lai2022PRL}, to name a few. Nevertheless, due to the weak COM coupling strength and the typical rapid decoherence of mechanical oscillators, it still remains a tough challenge to prepare highly entangled states based on COM system. To avoid such difficulties, various theoretical approaches have been developed by exploiting synthetic gauge field~\cite{Lai2022PRL,Liu2023SCMPA}, reservoir engineering~\cite{Wang2013PRL,Yang2015PRA,Sossa2023PRA}, dark-mode or feedback control~\cite{Lai2022PRR,Huang2022PRA2,Li2017PRA}, photon counting\cite{Ho2018PRL}, dynamical modulation~\cite{Wang2016PRA,Hu2019PRA,Huang2018PRA}, and optical nonreciprocity~\cite{Jiao2018PRL,Jiao2022PRAPP,Liu2022CPB}.

\begin{figure*}[htbp]
\centering
\includegraphics[width=0.9\textwidth]{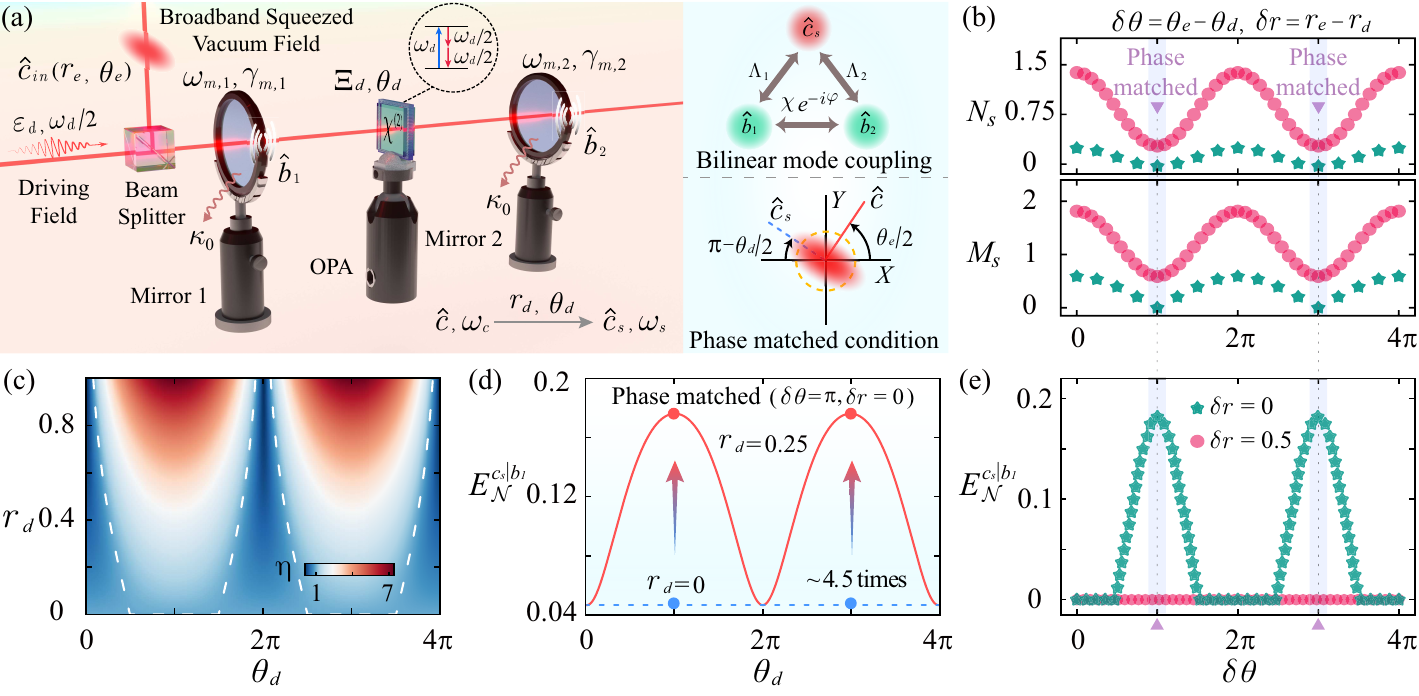}
\caption{\label{Fig_schematic} Enhancing light-mirror interactions and simultaneously suppressing light-reservoir interaction in squeezed COM systems. (a) Schematic diagram and interaction picture of a squeezed COM system consisting of an FP cavity and a nonlinear $\chi^{(2)}$ medium. The driving fields of this system include a coherent pump laser with amplitude $\varepsilon_{d}$ and frequency $\omega_{d}/2$, and a broadband squeezed laser with squeezing degree $r_{e}$ and reference phase $\theta_{e}$ around central frequency $\omega_{c}$. (b) The effective optical thermal noise $N_{s}$, the effective two-photon correlation strength $M_{s}$ and (e) the logarithmic negativity $E_{\mathcal{N}}^{\,\textit{c}_{\textit{s}}|\textit{b}_{\textit{1}}}$ versus the phase difference $\delta\theta$ for different values of $\delta r$. (c) Density plot of the enhancement factor $\eta$ of the COM interaction as a function of the squeezing degree $r_{d}$ and the squeezing reference angle $\theta_{d}$. (d) The logarithmic negativity $E_{\mathcal{N}}^{\,\textit{c}_{\textit{s}}|\textit{b}_{\textit{1}}}$ versus the squeezing reference angle $\theta_{d}$ under phase-matched condition achieving at $\delta r=0$ and $\delta\theta=\pm n\pi~(n=1,3,5,\ldots)$. The parameters are chosen as follows: $\omega_{m,j}/\omega_{m}=\textrm{1}$, $\kappa/\omega_{m}=\textrm{0.9}$, $\gamma_{m,j}/\omega_{m}=\textrm{10}^{-\textrm{5}}$, $g_{j}/\omega_{m}=\textrm{0.2}$, $\chi/\omega_{m}=\textrm{0.1}$, $\bar{n}_{m,j}=100$.}
\end{figure*}

In parallel, squeezed optical devices, which combine the techniques of OPA and reservoir engineering, open up a new route to enhance the light-matter interactions while suppress the input quantum noises, which boosts the achievement of a wealth of quantum phenomena. For example, several pioneering works on squeezed optical devices have contributed to the demonstration of photon blockade~\cite{Lu2015PRL}, entanglement between two atoms~\cite{Qin2018PRL}, long-lived or entangled cat states~\cite{Qin2021PRL,Chen2021PRL}, and optical squeezing beyond $\textrm{3\,dB}$ limit~\cite{Qin2022PRL}. Particularly, L\"{u} \textit{et al.} has predicted for the first time that by using a squeezed COM system, a single-photon quantum process can be observed even when the system is originally prepared in the weak COM coupling regime~\cite{Lu2015PRL}, which is owing to the exponential enhancement of the effective COM interaction in this system. Notably, such a strategy has also been utilized to realize nonreciprocal photon blockade~\cite{Tang2022PRL,Wang2023OE,Shen2023PRA} and nonreciprocal magnon laser~\cite{Huang2022OL} very recently. Besides, similar advances have also been achieved in making squeezed acoustic devices by employing quadratic COM coupling or mechanical parametric drive~\cite{Lemonde2016NC,Li2020PRL,Hei2023PRL}, which enables a variety of quantum effects, such as quantum phase transition~\cite{Lu2018PRAPP}, quantum chaos~\cite{Zhu2019PRA}, spin squeezing~\cite{Zhang2021PRA}, photon blockade~\cite{Wang2022PRAPP}, and coherent state transfer~\cite{Wang2023SCMPA}. However, as far as we know, the possibility of exploiting such powerful platform to create and manipulate highly entangled states with respect to massive mechanical oscillators, as well as protect the states from input noises, has not yet been reported up to now.

In this paper, we present how to achieve enhanced and controllable quantum entanglement by using a squeezed COM system. Here we consider a broadband squeezed laser field to be applied to a hybrid COM system consisting of a Fabry-P\'{e}rot (FP) cavity and an OPA, which we refer to as the squeezed COM system. We show that the interplay of the OPA and the squeezed laser field allows us to regulate the strength of both COM interaction and light-reservoir interaction simultaneously. This ability of the squeezed COM system plays a key role in the enhancement and manipulation of quantum entanglement. Interestingly, it is found that by tuning the squeezing parameters of the OPA, the degree of light-mirror, mirror-mirror and full tripartite entanglement, as well as the associated single-mode and two-mode quadrature squeezing, can be well regulated in a coherent and tunable way. Moreover, when further choosing proper squeezing parameters of the injected squeezed laser field to suppress the optical input noises, the COM entanglement, especially the tripartite one, can be considerably enhanced. Our proposal, providing a promising method to improve the performance of COM devices with multiple mechanical mirrors, can find its potential applications in diverse areas, such as building phononic quantum network~\cite{XuRMP2020,Kimble2008Nature,Kuzyk2018PRX}, synchronizing mechanical motions~\cite{Li2016OE,Sheng2020PRL,Zhu2021PR}, and realizing dual-comb spectroscopy~\cite{Ren2020NC}. These findings will benefit not only for the exploration of macroscopic quantum phenomena, but also for the achievement of various entanglement-enabled quantum technologies, such as optomechanical quantum information processing~\cite{Barzanjeh2022NP,Stannigel2012PRL} and weak-force sensing~\cite{Fleury2015NC,Zhao2020SCMPA}.

This paper is structured as follows. In Sec.\,\ref{secII}, we introduce the theoretical model of the squeezed COM system and derive the linearized Hamiltonian and the effective master equation of this system. In Sec.\,\ref{secIII}, we first calculate the covariance matrix of this system and then evaluate the entanglement measures. Based on this numerical simulations, we explore the generation and manipulation of the light-mirror, mirror-mirror and full tripartite entanglement in this squeezed COM system. In Sec.\,\ref{secIV}, a summary is given.

\section{Theoretical model and dynamics of a squeezed COM system}\label{secII}

In this paper, we propose to generate, manipulate, and enhance quantum entanglement in a tunable and coherent way with a squeezed COM setup. As shown in Fig.\,\ref{Fig_schematic}(a), we consider a hybrid COM system consisting of an FP cavity and a nonlinear $\chi^{(2)}$ medium. The FP cavity supports one single optical mode with resonance frequency $\omega_{c}$ and two mechanical vibrating modes with fundamental frequencies $\omega_{m,1}$ and $\omega_{m,2}$, respectively. The nonlinear $\chi^{(2)}$ medium can serve as an OPA that is pumped with frequency $\omega_{d}$, amplitude $\Xi_{d}$, and phase $\theta_{d}$. In a frame rotating with driving frequency $\omega_{d}/2$, the Hamiltonian of this system can be written as (setting $\hbar=1$)
\begin{align}
H=&H_{\textrm{c}}+H_{\textrm{m}}+H_{\textrm{om}}+H_{\textrm{dr}},
\notag \\
H_{\textrm{c}}=&\Delta_{c}\hat{c}^{\dagger}\hat{c}+\Xi_{d}(e^{-i\theta_{d}}\hat{c}^{\dagger2}+e^{i\theta_{d}}\hat{c}^{2}),
\notag \\
H_{\textrm{m}}=&\omega_{m,1}\hat{b}_{1}^{\dagger}\hat{b}_{1}+\omega_{m,2}\hat{b}_{2}^{\dagger}\hat{b}_{2}
+\chi(e^{-i\varphi}\hat{b}_{1}^{\dagger}\hat{b}_{2}+e^{i\varphi}\hat{b}_{2}^{\dagger}\hat{b}_{1}), \notag \\
H_{\textrm{om}}=&-g_{1}\hat{c}^{\dagger}\hat{c}(\hat{b}_{1}^{\dagger}+\hat{b}_{1})-g_{2}\hat{c}^{\dagger}\hat{c}(\hat{b}_{2}^{\dagger}+\hat{b}_{2}),
\notag \\
H_{\textrm{dr}}=&F_{1}(\hat{b}_{1}^{\dagger}+\hat{b}_{1})+F_{2}(\hat{b}_{2}^{\dagger}+\hat{b}_{2})
+i\varepsilon_{d}(\hat{c}^{\dagger}-\hat{c}),\label{eq_Ham1}
\end{align}
where $\hat{c}$ ($\hat{c}^{\dagger}$) and $\hat{b}_{j}$ ($\hat{b}_{j}^{\dagger}$) are the annihilation (creation) operators of the cavity mode and the $j$th $(j=1,2)$ mechanical mode, respectively, and $\Delta_{c}=\omega_{c}-\omega_{d}/2$ is the cavity-pump detuning. The first term $H_{\textrm{c}}$ in Eq.\,(\ref{eq_Ham1}) is the Hamiltonian of the cavity mode, in which the quadratic terms describe the nonlinear optical interaction induced by OPA. The second term $H_{\textrm{m}}$ denotes the Hamiltonian of the mechanical mode, with $\chi$ the coupling strength of the phonon-hopping interaction and $\varphi$ the corresponding modulation phase. The third term $H_{\textrm{om}}$ describes Hamiltonian for the optical-radiation-pressure mediated interaction between the cavity mode and the two mechanical modes, and  $g_{j}=(\omega_{c,j}/R_{j})\sqrt{\hbar/m_{j}\omega_{m,j}}$ denotes the single-photon COM coupling strength, with $R_{j}$ and $m_{j}$ being the cavity length and the effective mass of the $j$th mechanical resonator, respectively. The last term $H_{\textrm{dr}}$ is the Hamiltonian of the driving field, which includes two constant mechanical driving forces $F_{j}$ and a coherent optical pump field. The constant mechanical driving forces $F_{j}$ are applied to cancel the force induced by the parametric amplification (see detailed discussions below). The amplitude of the optical pump field is given by $|\varepsilon_{d}|=\sqrt{2\kappa P_{d}/\hbar\omega_{d}}$, where $P_{d}$ is the input laser power and $\kappa$ is the cavity decay rate.

In our proposed scheme, a broadband squeezed laser field $\hat{c}_{in}$, with squeezing degree $r_{e}$ and reference phase $\theta_{e}$ around central frequency $\omega_{c}$, is also introduced to act as the squeezed-vacuum reservoir that is linearly coupled to the FP cavity. For the two mechanical mirrors, they are assumed to be coupled with two independent thermal reservoirs at same bath temperature $T$. Considering such system-bath interactions for the optical and mechanical modes, the dynamics of the total system is governed by the following Born-Markovian master equation
\begin{align}
\dot{\rho}=&i\big[\rho,H\big]
+\frac{\kappa}{2}(N+1)\mathcal{D}[\hat{c}]\rho
+\frac{\kappa}{2}N\mathcal{D}[\hat{c}^{\dagger}]\rho
\notag\\
&-\frac{\kappa}{2}M\mathcal{G}[\hat{c}]\rho
-\frac{\kappa}{2}M^{\ast}\mathcal{G}[\hat{c}^{\dagger}]\rho
\notag\\
&+\sum_{j=1,2}\dfrac{\gamma_{m,j}}{2}(\bar{n}_{m,j}+1)\mathcal{D}[\hat{b}_{j}]\rho
+\dfrac{\gamma_{m,j}}{2}\bar{n}_{m,j}\mathcal{D}[\hat{b}_{j}^{\dagger}]\rho, \label{eq_meq}
\end{align}
where
\begin{align}
&\mathcal{D}[\hat{o}]\rho=2\hat{o}\rho\hat{o}^{\dagger}-\hat{o}^{\dagger}\hat{o}\rho-\rho\hat{o}^{\dagger}\hat{o},
\notag \\
&\mathcal{G}[\hat{o}]\rho=2\hat{o}\rho\hat{o}-\hat{o}\hat{o}\rho-\rho\hat{o}\hat{o}~~~(\hat{o}=\hat{c},~\hat{b}_{j}),
\end{align}
are the Lindblad operators, $[\cdot,\cdot]$ denotes the commutator, and $\gamma_{m,j}$ is the mechanical damping rate. $N=\sinh^{2}(r_{e})$ and $M=\cosh(r_{e})\sinh(r_{e})e^{i\theta_{e}}$ describe the optical thermal noise and the strength of the two-photon correlation caused by the squeezed-vacuum reservoir, respectively. $\bar{n}_{m,j}=1/[\exp(\omega_{m,j}/k_{B}T)-1]$ represents the mean thermal phonon excitation number of the $j$th mechanical modes, with $k_{B}$ the Boltzmann constant. Note that to obtain the master equation (\ref{eq_meq}), we have assumed each mirror has a high quality factor, namely, $Q_{m,j}\equiv\omega_{m,j}/\gamma_{m,j}\gg1$, which ensures that the system-bath interaction for the mechanical mode can be described by a Markovian process.

As discussed in detail in Ref.\,\cite{Lu2015PRL}, the light-mirror interaction can be exponentially enhanced with the facility of the generation of squeezed cavity field enabled by OPA. Here, in order to introduce the squeezed cavity field in our system, one can perform the Bogoliubov transformation with a unitary operator, $S(\eta_{d})=\exp[(-\eta_{d}\hat{c}^{\dagger2}+\eta_{d}^{\ast}\hat{c}^{2})/2]$, i.e.,
\begin{align}
S^{\dagger}(\eta_{d})\hat{c}S(\eta_{d})&=\cosh(r_{d})\hat{c}_{s}-e^{-i\theta_{d}}\sinh(r_{d})\hat{c}_{s}^{\dagger},
\end{align}
where $\eta_{d}=r_{d}e^{-i\theta_{d}}$ is the complex squeezing parameter, with a squeezing strength $r_{d}=(1/4)\ln[(\Delta_{c}+2\Xi_{d})/(\Delta_{c}-2\Xi_{d})]$ and a squeezing reference angle $\theta_{d}$. By applying the above Bogoliubov transformation and dropping the constant terms, the effective Hamiltonian of the total system is derived as
\begin{align}
\tilde{H}\!=
&\omega_{s}\hat{c}_{s}^{\dagger}\hat{c}_{s}\!+\!\omega_{m,1}\hat{b}_{1}^{\dagger}\hat{b}_{1}\!+\!\omega_{m,2}\hat{b}_{2}^{\dagger}\hat{b}_{2}
\!+\!\chi(e^{-i\varphi}\hat{b}_{1}^{\dagger}\hat{b}_{2}\!+\!e^{i\varphi}\hat{b}_{2}^{\dagger}\hat{b}_{1})
\notag \\
&-\!\zeta_{s,1}\hat{c}_{s}^{\dagger}\hat{c}_{s}(\hat{b}_{1}^{\dagger}\!+\!\hat{b}_{1})\!+\!\dfrac{\zeta_{p,1}}{2}(e^{-i\theta_{d}}\hat{c}_{s}^{\dagger2}\!+\!e^{i\theta_{d}}\hat{c}_{s}^{2})(\hat{b}_{1}^{\dagger}\!+\!\hat{b}_{1})
\notag \\
&-\!\zeta_{s,2}\hat{c}_{s}^{\dagger}\hat{c}_{s}(\hat{b}_{2}^{\dagger}\!+\!\hat{b}_{2})\!+\!\dfrac{\zeta_{p,2}}{2}(e^{-i\theta_{d}}\hat{c}_{s}^{\dagger2}\!+\!e^{i\theta_{d}}\hat{c}_{s}^{2})(\hat{b}_{2}^{\dagger}\!+\!\hat{b}_{2})
\notag \\
&+\!i\varepsilon_{d}\!\cosh(r_{d})(\hat{c}_{s}^{\dagger}\!-\!\hat{c}_{s})
\!+\!i\varepsilon_{d}\!\sinh(r_{d})(e^{-i\theta_{d}}\hat{c}_{s}^{\dagger}\!-\!e^{i\theta_{d}}\hat{c}_{s}).
\label{eq_H2}
\end{align}
Here, $\omega_{s}=(\Delta_{c}-2\Xi_{d})\exp(2r_{d})$ is referred to as the effective resonance frequency of the squeezed cavity mode, and we have chosen $F_{j}=g_{j}\sinh^{2}(r_{d})$ to cancel the constant mechanical force induced by parametric amplification. Moreover, the effective COM coupling strength of the $j$th mechanical mode induced by the radiation pressure and the parametric amplification can be respectively expressed as
\begin{align}
&\zeta_{s,j}=g_{j}\cosh(2r_{d})=\dfrac{g_{j}\Delta_{c}}{\sqrt{\Delta_{c}^{2}-4\Xi_{d}^{2}}},\notag\\
&\zeta_{p,j}=g_{j}\sinh(2r_{d})=\dfrac{2g_{j}G}{\sqrt{\Delta_{c}^{2}-4\Xi_{d}^{2}}}.
\end{align}
Correspondingly, the effective master equation under the Bogoliubov transformation can be derived as
\begin{align}
\dot{\tilde{\rho}}
=&i\big[\tilde{\rho},\tilde{H}\big]+\frac{\kappa}{2}(N_{s}+1)\mathcal{D}[\hat{c}_{s}]\tilde{\rho}
+\frac{\kappa}{2}N_{s}\mathcal{D}[\hat{c}_{s}^{\dagger}]\tilde{\rho}
\notag \\
&-\frac{\kappa}{2}M_{s}\mathcal{G}[\hat{c}_{s}]\tilde{\rho}
-\frac{\kappa}{2}M_{s}^{\ast}\mathcal{G}[\hat{c}_{s}^{\dagger}]\tilde{\rho}
\notag \\
&+\sum_{j=1,2}\dfrac{\gamma_{m,j}}{2}(\bar{n}_{m,j}+1)\mathcal{D}[\hat{b}_{j}]\tilde{\rho}+\dfrac{\gamma_{m,j}}{2}\bar{n}_{m,j}\mathcal{D}[\hat{b}_{j}^{\dagger}]\tilde{\rho},
\label{eq_meq2}
\end{align}
where
\begin{align}
N_{s}=&\sinh^{2}(r_{d})\cosh^{2}(r_{e})+\cosh^{2}(r_{d})\sinh^{2}(r_{e})
\notag \\
&+\dfrac{1}{2}\cos(\delta\theta)\sinh(2r_{d})\sinh(2r_{e}),\notag \\
M_{s}=&e^{i\theta_{d}}[\cosh(r_{d})\cosh(r_{e})+e^{-i\delta\theta}\sinh(r_{d})\sinh(r_{e})] \notag \\
&\times[\sinh(r_{d})\cosh(r_{e})+e^{i\delta\theta}\cosh(r_{d})\sinh(r_{e})] \label{eq_nsms}
\end{align}
are the effective optical thermal noise and the effective two-photon correlation strength, respectively. Here $\delta\theta\equiv\theta_{e}-\theta_{d}$ is the phase difference between the squeezed intracavity field and the squeezed-vacuum reservoir, and $\tilde{\rho}=S^{\dagger}(\eta_{d})\rho S(\eta_{d})$ is an effective density operator of the total system. The detailed derivation process of the effective Hamiltonian (\ref{eq_H2}) and the associated master equation (\ref{eq_meq2}) is reported in the Supporting Information. From Eq.\,(\ref{eq_nsms}), one can find that by choosing an equal squeezing strength of both the intracavity mode and the squeezed-vacuum reservoir, i.e., $r_{d}=r_{e}=r$, $N_{s}$ and $M_{s}$ can be simply reduced to
\begin{align}
\nonumber
N_{s}=&\dfrac{1}{2}\sinh^{2}(2r)[1+\cos(\delta\theta)],\\
M_{s}=&\dfrac{1}{2}e^{i\theta_{d}}\sinh(2r)(1+e^{i\delta\theta})
[\cosh^{2}(r)+e^{-i\delta\theta}\sinh^{2}(r)].
\label{eq_nsms2}
\end{align}
Equation\,(\ref{eq_nsms2}) implies that when the phase-matched condition between the squeezed intracavity field and the squeezed-vacuum reservoir is fulfilled by setting $\delta r\equiv r_{e}-r_{d}=0$ and $\delta\theta=\pm n\pi~(n=1,3,5,\ldots)$, the effective optical thermal noise $N_{s}$ and the effective two-photon-correlation strength $M_{s}$ can thus be completely eliminated, that is, $N_{s}=M_{s}=0$. In contrast, for other values of $\delta r$ and $\delta\theta$, $N_{s}$ and $M_{s}$ cannot be reduced to zero, which is corresponding to the phase-mismatched regime. Figure\,\ref{Fig_schematic}(b) shows the dependence of the optical thermal noise $N_{s}$ and the two-photon correlation strength $M_{s}$ on the squeezing parameters $\delta r$ and $\delta\theta$. This numerical result is consistent with the qualitative discussion, which is also reminiscent of those similar findings for squeezed quantum systems in the preceding studies~\cite{Lu2015PRL,Qin2018PRL,Qin2021PRL,Chen2021PRL,Qin2022PRL}. These results mean that in the basis of the squeezed cavity field $\hat{c}_{s}$ and under the phase-matched condition, i.e., $N_{s}=M_{s}=0$, the squeezed-vacuum reservoir originally coupled to the cavity field $\hat{c}$ becomes an effective vacuum reservoir [see Fig.\,\ref{Fig_schematic}(a) for more details]. Instead, for the phase-mismatched regime, the system will still be coupled to the squeezed-vacuum reservoir since the corresponding optical noise $N_{s}$ and $M_{s}$ is not equal to zero. Therefore, in order to enhance the COM coupling rate and simultaneously suppress the effective optical input noises, an essential procedure is to couple the system with a squeezed-vacuum reservoir and then properly choose the squeezing parameters to satisfy the phase-matched condition. Such squeezed COM systems with the phase-matched squeezed-vacuum reservoir have many advantages in the generation and manipulation of entanglement between light and mirrors, which will be further analyzed in the following discussions.

The Hamiltonian involved in master equation (\ref{eq_meq2}) includes the terms of nonlinear COM interactions between the optical field and the mechanical mirrors. In this case, the dynamics of this system is difficult to be directly solved. In order to deal with this problem, one can linearize the Hamiltonian by expanding each operator as a sum of its steady-state mean value and a small quantum fluctuation around it under the condition of strong optical driving, i.e.,
\begin{align}
\hat{c}_{s}=\bar{c}_{s}+\delta\hat{c}_{s},~~
\hat{b}_{j}=\bar{b}_{j}+\delta\hat{b}_{j}. \label{eq_ss1}
\end{align}
By using the master equation (\ref{eq_meq2}), one can obtain the steady-state mean value of the optical field and the mechanical oscillators as
\begin{align}
\bar{c}_{s}=&-\dfrac{(i\Delta_{s}-\dfrac{\kappa}{2})A_{1}+iA_{2}\beta_{p}}
{(\Delta_{s}^2+\dfrac{\kappa^{2}}{4})-\beta_{p}^{2}}\varepsilon_{d},
\notag \\
\bar{b}_{1}=&\dfrac{i\zeta_{s,1}|\bar{c}_{s}|^{2}
-i\dfrac{\zeta_{p,1}}{2}\alpha_{s}-i\chi e^{-i\varphi}\bar{b}_{2}}{i\omega_{m,1}+\dfrac{\gamma_{m,1}}{2}},
\notag \\
\bar{b}_{2}=&\dfrac{i\zeta_{s,2}|\bar{c}_{s}|^{2}
-i\dfrac{\zeta_{p,2}}{2}\alpha_{s}-i\chi e^{i\varphi}\bar{b}_{1}}{i\omega_{m,2}+\dfrac{\gamma_{m,2}}{2}}, \label{eq_ss2}
\end{align}
with
\begin{align}
\Delta_{s} &= \omega_{s}-\beta_{s}, \notag \\
\alpha_{s} &= e^{-i\theta_{d}}\bar{c}_{s}^{\ast2}+e^{i\theta_{d}}\bar{c}_{s}^{2}, \notag \\
\beta_{s} &= \zeta_{s,1}(\bar{b}_{1}^{\ast}+\bar{b}_{1})+\zeta_{s,2}
(\bar{b}_{2}^{\ast}+\bar{b}_{2}),\notag \\
\beta_{p} &= \zeta_{p,1}(\bar{b}_{1}^{\ast}+\bar{b}_{1})+\zeta_{p,2}(\bar{b}_{2}^{\ast}+\bar{b}_{2}), \notag \\
A_{1} &= \cosh(r_{d})+\sinh(r_{d})e^{-i\theta_{d}}, \notag \\
A_{2} &= \cosh(r_{d})e^{-i\theta_{d}}+\sinh(r_{d}).
\end{align}
Then, by using Eqs.\,(\ref{eq_ss1}) and (\ref{eq_ss2}), one can directly derive the linearized Hamiltonian and the associated master equation as
\begin{align}
\tilde{H}_{lin}
=&\Delta_{s}\hat{c}_{s}^{\dagger}\hat{c}_{s}+\omega_{m,1}\hat{b}_{1}^{\dagger}\hat{b}_{1}+\omega_{m,2}\hat{b}_{2}^{\dagger}\hat{b}_{2}
\notag \\
&+\!\dfrac{1}{2}\beta_{p}(e^{-i\theta_{d}}\hat{c}_{s}^{\dagger2}\!+\!e^{i\theta_{d}}\hat{c}_{s}^{2})
\!+\!\chi(e^{-i\varphi}\hat{b}_{1}^{\dagger}\hat{b}_{2}\!+\!e^{i\varphi}\hat{b}_{2}^{\dagger}\hat{b}_{1})
\notag \\
&-\!(\Lambda_{1}\hat{c}_{s}^{\dagger}\!+\!\Lambda_{1}^{\ast}\hat{c}_{s})(\hat{b}_{1}^{\dagger}\!+\!\hat{b}_{1})
\!-\!(\Lambda_{2}\hat{c}_{s}^{\dagger}\!+\!\Lambda_{2}^{\ast}\hat{c}_{s})(\hat{b}_{2}^{\dagger}\!+\!\hat{b}_{2}),
\label{eq_Hlin}
\end{align}
and
\begin{align}
\dot{\tilde{\rho}}
=&i\big[\tilde{\rho},\tilde{H}_{lin}\big]+\frac{\kappa}{2}(N_{s}+1)\mathcal{D}[\hat{c}_{s}]\tilde{\rho}
+\frac{\kappa}{2}N_{s}\mathcal{D}[\hat{c}_{s}^{\dagger}]\tilde{\rho}
\notag \\
&-\frac{\kappa}{2}M_{s}\mathcal{G}[\hat{c}_{s}]\tilde{\rho}
-\frac{\kappa}{2}M_{s}^{\ast}\mathcal{G}[\hat{c}_{s}^{\dagger}]\tilde{\rho}
\notag \\
&+\sum_{j=1,2}\dfrac{\gamma_{m,j}}{2}(\bar{n}_{m,j}+1)\mathcal{D}[\hat{b}_{j}]\tilde{\rho}+\dfrac{\gamma_{m,j}}{2}\bar{n}_{m,j}\mathcal{D}[\hat{b}_{j}^{\dagger}]\tilde{\rho},
\label{eq_meq3}
\end{align}
where
\begin{align}
\Lambda_{j}=G_{j}\cosh(2r_{d})-G_{j}^{\ast}\sinh(2r_{d})e^{-i\theta_{d}} \label{eq_coupling}
\end{align}
is the effective COM coupling rate, with $G_{j}=g_{j}\bar{c}_{s}$. For notational convenience, we have neglected the symbol $``\delta"$ in the expression of quantum fluctuation operators in Eqs.\,(\ref{eq_Hlin}) and (\ref{eq_meq3}). We also note that in the weak COM coupling regime, the steady-state mean value of the optical mode is much larger than those of the mechanical modes, i.e., $|\bar{c}_{s}|\gg|\bar{b}_{j}|$. Therefore, under this condition, the terms $\hat{c}_{s}^{2}$ and $\hat{c}_{s}^{\dagger2}$, whose coefficient $\beta_{p}$ is proportional to $\bar{b}_{j}$, can be safely ignored in Hamiltonian\,(\ref{eq_Hlin}). Then, the Hamiltonian\,(\ref{eq_Hlin}) can be reduced to
\begin{align}
\tilde{H}_{lin}
\simeq&\Delta_{s}\hat{c}_{s}^{\dagger}\hat{c}_{s}+\omega_{m,1}\hat{b}_{1}^{\dagger}\hat{b}_{1}+\omega_{m,2}\hat{b}_{2}^{\dagger}\hat{b}_{2}
+\!(\nu\hat{b}_{1}^{\dagger}\hat{b}_{2}\!+\!\nu^{\ast}\hat{b}_{2}^{\dagger}\hat{b}_{1})
\notag \\
&-(\Lambda_{1}\hat{c}_{s}^{\dagger}\!+\!\Lambda_{1}^{\ast}\hat{c}_{s})(\hat{b}_{1}^{\dagger}\!+\!\hat{b}_{1})
-\!(\Lambda_{2}\hat{c}_{s}^{\dagger}\!+\!\Lambda_{2}^{\ast}\hat{c}_{s})(\hat{b}_{2}^{\dagger}\!+\!\hat{b}_{2}), \label{eq_Hlin2}
\end{align}
where $\nu=\chi e^{-i\varphi}$. In our following numerical calculations, for ensuring the validity of Eq.\,(\ref{eq_Hlin2}), the parameters have been strictly restricted to satisfy the condition of weak COM coupling.

Moreover, as shown in Ref.\,\cite{Lu2015PRL}, the ability of squeezed COM system to regulate its intracavity field intensity by adjusting the squeezing parameters provides a feasible and efficient way to enhance the strength of light-mirror interaction. To clearly see this, an enhancement factor $\eta_{j}$, quantifying the amount by which the effective COM coupling strength is enhanced in comparison with that of the conventional COM system, is defined as
\begin{align}
\eta_{j} = |\Lambda_{j}/G_{j}|.
\end{align}
Under the condition of equal single-photon COM coupling strength, i.e., $g_{1}/g_{2}=1$, we have $\eta_{1}=\eta_{2}=\eta$. In Fig.\,\ref{Fig_schematic}(c), $\eta$ is plotted as a function of the squeezing strength $r_{d}$ and the squeezing reference angle $\theta_{d}$. Here, $G_{j}$ has been assumed to be real, which can be achieved by choosing a suitable phase of $\bar{c}_{s}$ through tuning the phase reference of the cavity field. Interestingly, it is seen that by properly adjusting the squeezing parameters of the squeezed intracavity mode, e.g., when choosing $r_{d}=1$ and $\theta_{d}=\pi$ or $3\pi$, the maximum value of $\eta$ can reach up to $\sim7$, indicating that the effective COM coupling strength $\Lambda_{j}$ are improved by $7$ times in comparison with that of the conventional COM system.

\section{Enhancing bipartite and tripartite entanglement under the phase-matched condition}\label{secIII}

As discussed above, we derive an effective theoretical model for the squeezed COM system. In this section, based on this model, we study how to generate, manipulate, and even enhance the bipartite light-mirror and mirror-mirror entanglement, as well as the full tripartite entanglement. For this purpose, we shall first calculate the covariance matrix (CM) for the squeezed COM system, by which one can quantify the bipartite and tripartite entanglement through introducing proper entanglement measures, e.g., the logarithmic negativity and the minimum residual contangle. Then, based on numerical calculations of such measures, we will qualitatively and quantitatively discuss some specific cases with regard to the coherent generation and manipulation of entanglement in squeezed COM systems, and show its advantages on the achievement of enhanced bipartite and tripartite entanglement under the phase-matched condition.

\begin{figure*}[htbp]
\centering
\includegraphics[width=0.9\textwidth]{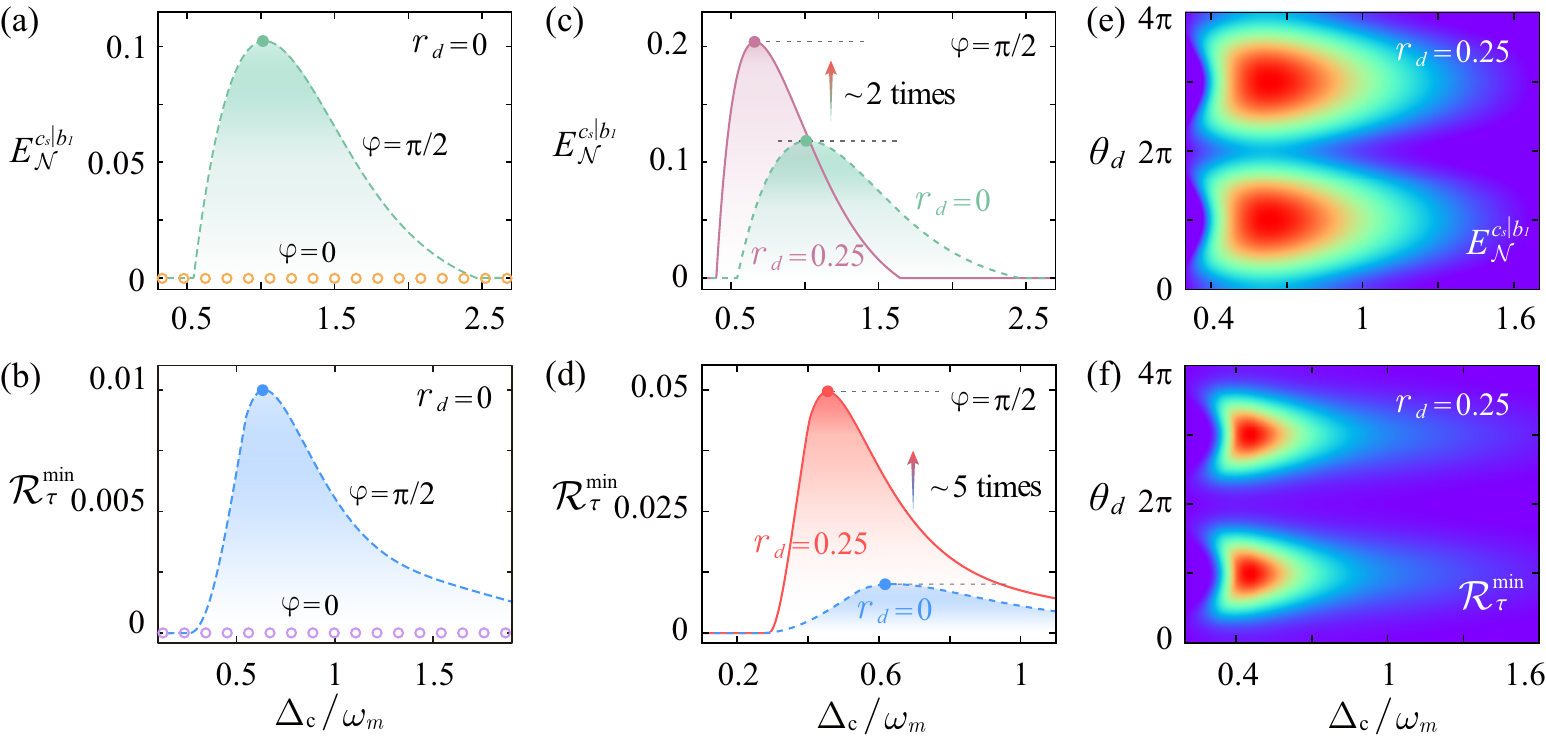}
\caption{\label{Fig_omen} Enhancement of bipartite light-mirror entanglement and full tripartite entanglement under the phase-matched condition achieving at $\delta r=0$ and $\delta\theta=\pm n\pi~(n=1,3,5,\ldots)$. (a, c) The logarithmic negativity $E_{\mathcal{N}}^{\,\textit{c}_{\textit{s}}|\textit{b}_{\textit{1}}}$ and (b, d) the minimum residual contangle $\mathcal{R}^{\textrm{min}}_{\tau}$ versus the scaled optical detuning $\Delta_{c}/\omega_{m}$ for different values of squeezing strength $r_{d}$, with squeezing reference angle $\theta_{d}=\pi$. For $r_{d}=0$, $E_{\mathcal{N}}^{\,\textit{c}_{\textit{s}}|\textit{b}_{\textit{1}}}$ and $\mathcal{R}^{\textrm{min}}_{\tau}$ can achieve their maximum values when breaking the dark mode effect at $\varphi=\pi/2$. For $r_{d}=0.25$, with respect to the case of DMB at $\varphi=\pi/2$, the maximum values of $E_{\mathcal{N}}^{\,\textit{c}_{\textit{s}}|\textit{b}_{\textit{1}}}$ and $\mathcal{R}^{\textrm{min}}_{\tau}$ are further enhanced by about $2$ and $5$ times, respectively. (e, f) Density plot of $E_{\mathcal{N}}^{\,\textit{c}_{\textit{s}}|\textit{b}_{\textit{1}}}$ and $\mathcal{R}^{\textrm{min}}_{\tau}$ as a function of the scaled optical detuning $\Delta_{c}/\omega_{m}$ and the squeezing reference angle $\theta_{d}$, with squeezing strength $r_{d}=0.25$. Other parameters are chosen as those in Fig.\,\ref{Fig_schematic}.}
\end{figure*}

\subsection{Covariance matrix and quantitative entanglement measures for squeezed COM systems}

In our proposed scheme, due to the linearized dynamics and the Gaussian nature of the optical and mechanical input noises, the steady state of the squeezed COM system, independently of any initial condition, could eventually evolve into a tripartite zero-mean Gaussian state, whose statistic is commonly characterized by a Wigner function with multivariate normal distribution on phase space, i.e.,
\begin{align}
\textrm{W}(\psi)=\dfrac{\exp\left[-\dfrac{1}{2}\left(\psi^{\dagger}V^{-1}\psi\right)\right]}{\pi^{2}\sqrt{\det(V)}}.
\end{align}
Here $\psi=(\hat{X},\hat{Y},\hat{Q}_{1},\hat{P}_{1},\hat{Q}_{2},\hat{P}_{2})^{T}$ is the vector of optical and mechanical quadrature operators, with its components defined by
\begin{align}
&X=\dfrac{1}{\sqrt{2}}\left(\hat{c}_{s}^{\dagger}+\hat{c}_{s}\right),~~~Y=\dfrac{i}{\sqrt{2}}\left(\hat{c}_{s}^{\dagger}-\hat{c}_{s}\right),\notag\\
&Q_{j}=\dfrac{1}{\sqrt{2}}\left(\hat{b}_{j}^{\dagger}+\hat{b}_{j}\right),~~~P_{j}=\dfrac{i}{\sqrt{2}}\left(\hat{b}_{j}^{\dagger}-\hat{b}_{j}\right).
\label{eq_quad}
\end{align}
Also, $V$ denotes the steady-state CM of the squeezed COM system, and its matrix element is given by
\begin{align}
V_{kl}=\langle \psi_{k}\psi_{l}+\psi_{l}\psi_{k}\rangle/2,~(k,l=1,2,\ldots,6).
\label{eq_Vkl}
\end{align}
By grouping together the quantum correlation functions for bosonic operators in a vector, $X=(x_{1},x_{2},\ldots,x_{24})^{T}$, one can obtain its time evolution equation from Eq.\,(\ref{eq_meq3}), i.e.,
\begin{align}
\dfrac{d}{dt}X=M\cdot X+N, \label{eq_eom}
\end{align}
where
\begin{align}
N=\big[&\kappa N_{s},\kappa(N_{s}+1),\gamma_{m,1}\bar{n}_{m,1},\gamma_{m,1}(\bar{n}_{m,1}+1),
\notag \\
&\gamma_{m,2}\bar{n}_{m,2}, \gamma_{m,2}(\bar{n}_{m,2}+1),\kappa M_{s}^{\ast},\kappa M_{s},0,0,
\notag \\
&0,0,0,0,0,0,0,0,0,0,0,0,0,0\big]^{T},
\end{align}
is a vector involving the correlation functions for the input quantum noises. Here
the exact expression for the coefficient matrix $M$ and the detailed derivation process of the time evolution equation\,(\ref{eq_eom}) are too cumbersome, and, for convenience, we have reported them in the Supporting Information. Notably, by numerically solving Eq.\,(\ref{eq_eom}) and using the relations between bosonic operators and quadrature operators, one can directly obtain the steady-state CM $V$.

Regarding the verification of the fundamental entanglement property in our proposed scheme, we adopt here the logarithmic negativity $E_{\mathcal{N}}$ and the minimum residual contangle $\mathcal{R}^{\textrm{min}}_{\tau}$ as quantitative measures for certifying the bipartite and tripartite entanglement, respectively, which are defined based on specifying the positivity of the partial transpose of the states. Considering that both optical and mechanical modes of this squeezed COM system are continuous variables (CV), the logarithmic negativity $E_{\mathcal{N}}$ is defined as~\cite{Adesso2004pra}
\begin{align}
E_{\mathcal{N}}=&\max\,\big[0,-\ln(2\eta_{0}^{-})\big],
\label{En}
\end{align}
where $\eta_{0}^{-}\!\equiv\!2^{-1/2}\{\Sigma(V_{\mu\nu})-[\Sigma(V_{\mu\nu})^{2}-4\det\!V_{\mu\nu}]^{1/2}\}^{1/2}$ , with $\Sigma(V_{\mu\nu})\!\equiv\!\det\mathcal{A}_{\mu}+\det\mathcal{B}_{\nu}-2\det\mathcal{C}_{\mu\nu}$, is the minimum symplectic eigenvalue of the partial transpose of a reduced $4\times4$ CM $V_{\mu\nu}$. The indexes $\mu$ and $\nu$ label the two selected modes under consideration, and the reduced CM $V_{\mu\nu}$ contains the entries of $V$ associated with the corresponding bipartitions $\mu$ and $\nu$. By removing the rows and columns of the unwanted modes in $V$, the reduced CM $V_{\mu\nu}$ can be straightforwardly obtained, which has a $ 2\times2 $ block form given by
\begin{align}
V_{\mu\nu}=\left(
\begin{matrix}
\mathcal{A}_{\mu}&\mathcal{C}_{\mu\nu}\\
\mathcal{C}_{\mu\nu}^{\textit{T}}&\mathcal{B}_{\nu}
\end{matrix}
\right).\label{reducedCM}
\end{align}
Equation\,(\ref{En}) indicates that the selected bipartition $\mu$ and $\nu$ gets entangled if and only if $\eta_{0}^{-}<1/2$, which is equivalent to Simon's necessary and sufficient entanglement nonpositive partial transpose criterion (or the related Peres-Horodecki criterion) for certifying bipartite entanglement in Gaussian states~\cite{Simon2000PRL}.

Furthermore, the minimum residual contangle, $\mathcal{R}^{\textrm{min}}_{\tau}$, which provides a \textit{bona fide} quantification of CV tripartite entanglement, is commonly defined as
\begin{align}
\mathcal{R}^{\textrm{min}}_{\tau}=&\min_{(r,s,t)}\,\big[E^{r|st}_{\tau}-E^{r|s}_{\tau}-E^{r|t}_{\tau}\big],
\end{align}
where $(r,s,t)\in\{c_{s},b_{1},b_{2}\}$ denotes all the possible permutations of the three-mode indexes. $E_{\tau}^{u|v}$ is the contangle of subsystems of $u$ ($u$ contains one mode) and $v$ ($v$ contains one or two modes), which can be defined by a proper entanglement monotone, e.g., the squared logarithmic negativity. Based on Eq.\,(\ref{En}), the one-mode-vs-one-mode contangle $E^{r|s}_{\tau}$ and $E^{r|t}_{\tau}$ can be directly obtained by employing its definition, namely, $E_{\tau}\equiv[E_{\mathcal{N}}]^{2}$. However, when calculating the one-mode-vs-two-modes contangle $E^{r|st}_{\tau}$, one must alter the basic definition of Eq.\,(\ref{En}) by rewriting the definition of $\eta_{0}^{-}$ as given by $\eta_{0}^{-}\equiv\min\,[\textrm{eig}|i\Omega_{3}\tilde{V}_{r|st}|]$, where $\eta_{0}^{-}$ becomes the minimum symplectic eigenvalue of the partial transpose of a $6\times6$ CM $V$, with $\Omega_{3}=\oplus_{k=1}^{3}i\sigma_{y}$ and $\sigma_{y}$ the y-Pauli matrix. $\tilde{V}_{r|st}$ corresponds to the partial transpose of $V$, which connects to $V$ with the relation of $\tilde{V}_{r|st}=P_{r|st}VP_{r|st}$, and $P_{r|st}=\textrm{diag}(1,-1,1,1,1,1)$ is the partial transposition matrix. In terms of $E^{s|rt}_{\tau}$ and $E^{t|sr}_{\tau}$, the corresponding partial transpose matrices are given by $P_{s|rt}=\textrm{diag}(1,1,1,-1,1,1)$ and $P_{t|sr}=\textrm{diag}(1,1,1,1,1,-1)$, respectively. In addition, according to the Coffman-Kundu-Wootters monogamy inequality for quantum entanglement, we also note that the residual contangle is required to satisfy the following monogamy condition, i.e., $E^{r|st}_{\tau}-E^{r|s}_{\tau}-E^{r|t}_{\tau}\geq0$, which means that the bipartite entanglement between the partition $r$ and the remaining two partitions $st$ is never smaller than the sum of the $r|s$ and $r|t$ bipartite entanglements in the reduced states. As such, if there are nonzero values of the minimum residual contangle, i.e., $\mathcal{R}_{\tau}^{\textrm{min}}>0$, one can verify that the full tripartite entanglement is present for the CV system.

\subsection{Numerical verification of the bipartite and tripartite entanglement in squeezed COM systems}

In the previous sections, as demonstrated in Figs.\,\ref{Fig_schematic}(b) and \ref{Fig_schematic}(c), we have shown that the squeezed COM system can provide an efficient method to regulate both the light-mirror interaction and the light-reservoir interaction, which thus allows us to enhance the effective COM coupling strength while suppress the effective optical input noises by choosing proper squeezing parameters. Hereafter, we start to explore how to coherently generate and manipulate entanglement by harnessing this unique property of the squeezed COM system. For evaluating entanglement measures, we can numerically calculate the steady-state CM $V$ by solving Eq.\,(\ref{eq_eom}) through using the following scaled parameters: $\omega_{m,j}/\omega_{m}=\textrm{1}$, $\kappa/\omega_{m}=\textrm{0.9}$, $\gamma_{m,j}/\omega_{m}=\textrm{10}^{-\textrm{5}}$, $g_{j}/\omega_{m}=\textrm{0.2}$, $\chi/\omega_{m}=\textrm{0.1}$, $\bar{n}_{m,j}=\textrm{100}$. Here we assume the same parameters for two identical mechanical mirrors, which ensures that the two types of light-mirror entanglement ($E_{\mathcal{N}}^{\,\textit{c}_{\textit{s}}|\textit{b}_{\textit{1}}}$ and $E_{\mathcal{N}}^{\,\textit{c}_{\textit{s}}|\textit{b}_{\textit{2}}}$) in our proposed scheme have similar behaviors with the variation of the controlling parameters. Based on this assumption, for convenience, we only study the behavior of $E_{\mathcal{N}}^{\,\textit{c}_{\textit{s}}|\textit{b}_{\textit{1}}}$ as an example in the following discussion. Also, we note that the values of the squeezing parameters $r_{d}$ and $r_{e}$ have been chosen ranging from about $\textrm{0.01}$ to $\textrm{1}$ in such calculations, which corresponds to a ratio for $\Xi_{d}/\Delta_{c}$ estimated on the order of $\textrm{10}^{-\textrm{3}}\sim\textrm{10}^{-\textrm{1}}$. We stress that, for FP cavities, $\omega_{m}$ is typically $\textrm{10}^{-\textrm{3}}-\textrm{10}^{\textrm{3}}\,\textrm{MHz}$, and thus to satisfy the above condition for $\Xi_{d}/\Delta_{c}$ around the COM resonance region (i.e., $\Delta_{c}/\omega_{m}\simeq\textrm{1}$), it requires the order of $\Xi_{d}$ ranging from $\textrm{10}^{-\textrm{6}}$ to $\textrm{10}^{\textrm{2}}\,\textrm{MHz}$, which is feasible for current experimental techniques~\cite{Bruch2019Optica,Liu2023FO}.

\begin{figure*}[htbp]
\centering
\includegraphics[width=0.9\textwidth]{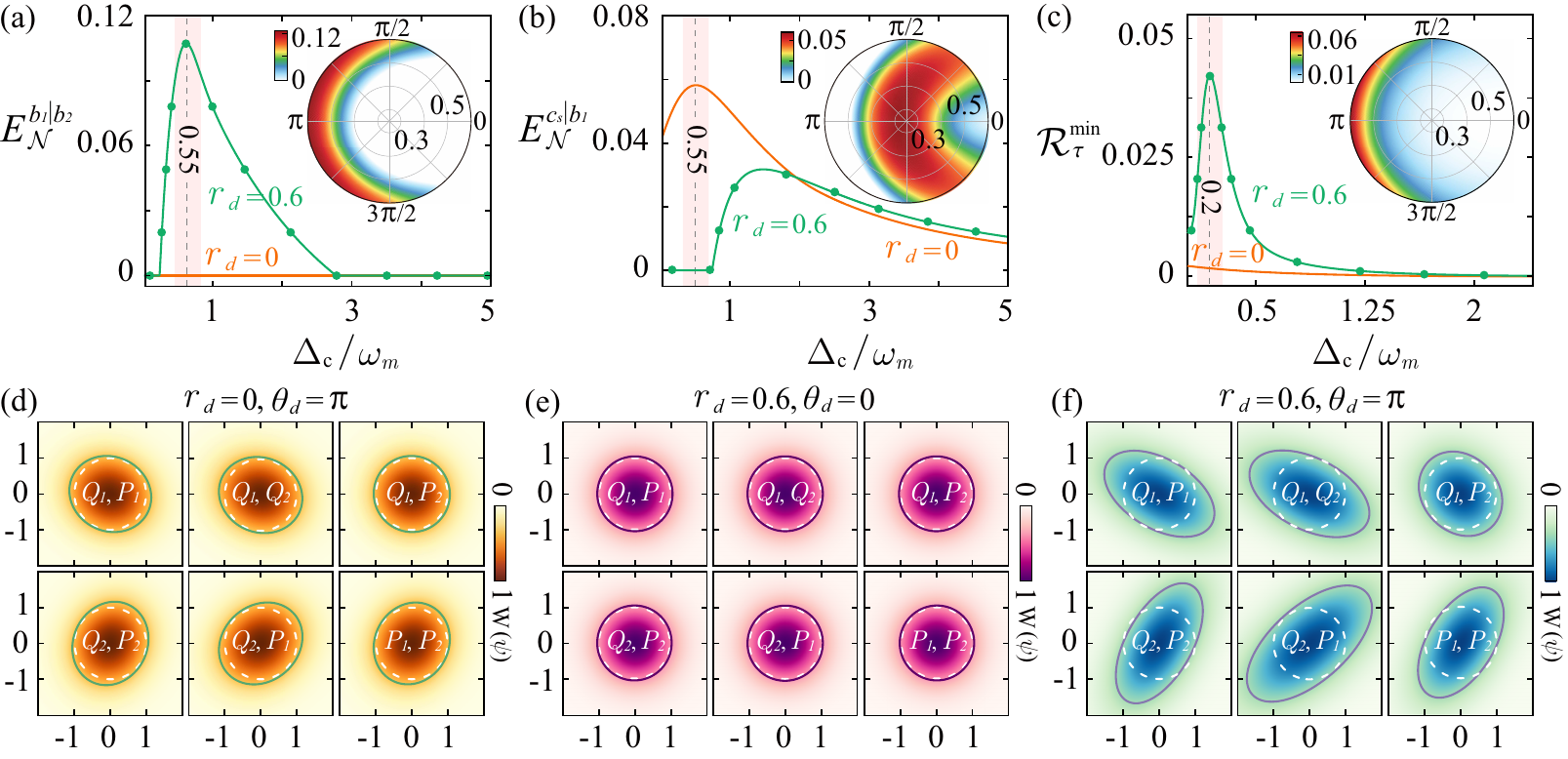}
\caption{\label{Fig_mmen} Entanglement and the associated quadrature squeezing with respect to the subsystem of two mechanical mirrors. The logarithmic negativity (a) $E_{\mathcal{N}}^{\,\textit{b}_{\textit{1}}|\textit{b}_{\textit{2}}}$, (b) $E_{\mathcal{N}}^{\,\textit{c}_{\textit{s}}|\textit{b}_{\textit{1}}}$, and (c) the minimum residual contangle $\mathcal{R}^{\textrm{min}}_{\tau}$ versus the scaled optical detuning $\Delta_{c}/\omega_{m}$ for different values of squeezing strength $r_{d}$, with squeezing reference angle $\theta_{d}=\pi$. The inserts show the dependence of $E_{\mathcal{N}}^{\,\textit{b}_{\textit{1}}|\textit{b}_{\textit{2}}}$, $E_{\mathcal{N}}^{\,\textit{c}_{\textit{s}}|\textit{b}_{\textit{1}}}$, and $\mathcal{R}^{\textrm{min}}_{\tau}$ on the squeezing strength $r_{d}$ and the squeezing reference angle $\theta_{d}$ at the optical detuning $\Delta_{c}/\omega_{m}=0.55$, $0.55$, $0.2$, respectively. (d)-(f) Reconstructed Wigner function $\textrm{W}(\psi)$ for the steady state of two mechanical modes at the optical detuning $\Delta_{c}/\omega_{m}=0.55$, with respect to the case for different values of squeezing strength $r_{d}$ and squeezing reference angle $\theta_{d}$, along with that of an ideal vacuum state for reference. Each panel corresponds to a characteristic projection of $\textrm{W}(\psi)$ on two-dimensional subspaces within $\psi \!=\! (\hat{X},\hat{Y},\hat{Q}_{1},\hat{P}_{1},\hat{Q}_{2},\hat{P}_{2})^{T}$, which is obtained by integrating out the other four quadratures in $\psi$. The solid (dashed) line indicates a drop by $1/\textrm{e}$ of the maximum value of $\textrm{W}(\psi)$ for the relevant steady (vacuum) state. The legends $\psi_{i}$, $\psi_{j}$ in each panel are corresponding to a correlation with $\psi_{i}$ along the $\textrm{x}$ axis and $\psi_{j}$ along the $\textrm{y}$ axis. The parameters are chosen as follows: $\gamma_{m,j}/\omega_{m}=0.2$, $g_{j}/\omega_{m}=0.1$, $\bar{n}_{m,j}=0.05$. Other parameters are chosen as those in Fig.\,\ref{Fig_schematic}.}
\end{figure*}

In Figs.\,\ref{Fig_schematic} and \ref{Fig_omen}, we first show how to regulate the behavior of bipartite and tripartite entanglement involving one mechanical mirror, which is achieved by adjusting the squeezing parameters of the squeezed intracavity mode and the injected squeezed laser field. Figure \ref{Fig_schematic}(d) shows the logarithmic negativity $E_{\mathcal{N}}^{\,\textit{c}_{\textit{s}}|\textit{b}_{\textit{1}}}$ as a function of the reference angle $\theta_{d}$ of the squeezed intracavity mode, with $\delta r=\textrm{0}$ and $\delta\theta=\pi$. One can intuitively see that in the absence of intracavity squeezing (i.e., $r_{d}=\textrm{0}$), $E_{\mathcal{N}}^{\,\textit{c}_{\textit{s}}|\textit{b}_{\textit{1}}}$ remains unchange with the variation of $\theta_{d}$, whose value always corresponds to a constant of about $\textrm{0.041}$. In contrast, when introducing the squeezed intracavity mode via OPA (i.e., $r_{d}=\textrm{0.25}$), $E_{\mathcal{N}}^{\,\textit{c}_{\textit{s}}|\textit{b}_{\textit{1}}}$ starts to oscillate back and forth periodically over $\theta_{d}$ with an approximate periodicity of $2\pi$. Moreover, the maximum value of $E_{\mathcal{N}}^{\,\textit{c}_{\textit{s}}|\textit{b}_{\textit{1}}}$ can reach up to about $\textrm{0.185}$ in this case, which is enhanced by $\textrm{4.5}$ times compared to that of the conventional COM system. This result indicates that our proposed scheme is poised to provide a promising platform for achieving the enhanced and controllable entangled states by coherently adjusting the reference angle of the squeezed intracavity mode, which we will discuss in detail later. The underlying physics of this phenomenon is due to the fact that the light-mirror entanglement is a monotonically increasing function of the effective COM coupling strength $\Lambda_{j}$, which, as shown in Fig.\,\ref{Fig_schematic}(c), are effectively regulated by the the squeezing parameters of the intracavity field. Then, for the squeezing parameters that can enhance the light-mirror interaction strength, it can correspondingly improve the light-mirror entanglement. However, it should be stressed that this procedure for entanglement manipulation and enhancement requires to be performed under the phase-matched condition, i.e., satisfying $\delta r=\textrm{0}$ and $\delta\theta=\pm n\pi~(n=\textrm{1},\textrm{3},\textrm{5},\ldots)$. To clearly see this, as demonstrated in Fig.\,\ref{Fig_schematic}(e), we plot the logarithmic negativity $E_{\mathcal{N}}^{\,\textit{c}_{\textit{s}}|\textit{b}_{\textit{1}}}$ as a function of the phase difference $\delta\theta$, with $r_{d}=\textrm{0.25}$ and $\theta_{d}=\textrm{0}$. It is obviously seen that $E_{\mathcal{N}}^{\,\textit{c}_{\textit{s}}|\textit{b}_{\textit{1}}}$ are highly peaked within a finite interval of values of $\delta\theta$ around $\delta\theta\simeq\pi$ or $\textrm{3}\pi$, showing that the light-mirror entanglement is sensitive to the variation of $\delta\theta$ and survives only around the phase-matched condition. Physically, this result can be understood as follows. The effective optical input noises $N_{s}$ and $M_{s}$ can only be suppressed completely when the phase-matched condition is fulfilled, whereas they would exponentially increase with other squeezing parameters, something we discussed in detail in Sec.\,\ref{secII}. On the other hand, when the optical input noises increase, the light-mirror entanglement commonly tends to be inhibited or even vanish. Therefore, it becomes a necessary condition for our proposed scheme to fulfill the phase-matched condition.

By assuming the fulfillment of the phase-matched condition, we further explore in detail how to achieve an enhancement of both bipartite light-mirror entanglement and full tripartite entanglement with respect to some specific squeezing parameters. In Figs.\,\ref{Fig_omen}(a)-\ref{Fig_omen}(d), we plot the logarithmic negativity $E_{\mathcal{N}}^{\,\textit{c}_{\textit{s}}|\textit{b}_{\textit{1}}}$ and the minimum residual contangle $\mathcal{R}^{\textrm{min}}_{\tau}$ as a function of the optical detuning $\Delta_{c}$ for different values of $r_{d}$, with $\theta_{d}=\pi$, $\delta r=\textrm{0}$, and $\delta\theta=\pi$. From Figs.\,\ref{Fig_omen}(a) and \ref{Fig_omen}(b), it is clearly seen that for the familiar case without any intracavity squeezing (i.e., $r_{d}=\textrm{0}$), to entangle an optical radiation field with mechanical mirrors in a three-mode COM system with bilinear cyclic couplings, one should first break the dark mode effect through choosing a proper phase $\varphi$ of the phonon-hopping interactions, which has already been revealed in some preceding studies~\cite{Huang2022PRA,Lai2022PRL,Huang2022PRA2,Huang2023PRA,Lai2022PRR}. Note that here, in the case of $\varphi=k\pi~(k=\textrm{0},\textrm{1},\textrm{2},\ldots)$, both bipartite light-mirror entanglement and full tripartite entanglement are absent, that is, $E_{\mathcal{N}}^{\,\textit{c}_{\textit{s}}|\textit{b}_{\textit{1}}}=\textrm{0}$ and $\mathcal{R}^{\textrm{min}}_{\tau}=\textrm{0}$, which corresponds to a situation of the dark-mode-unbroken (DMU) regime. In contrast, when choosing other values of $\varphi$, such bipartite and tripartite entanglement can simultaneously emerge and reach their maximum values at an optimal phase angle of $\varphi=\pi/\textrm{2}$, which corresponds to a situation of the dark-mode-broken (DMB) regime. From these results, it is found that the entanglement, especially the tripartite one, are severely restricted even in the DMB regime, which results from the limitation of weak light-mirror coupling strength of the conventional COM system. Interestingly, as shown in Figs.\,\ref{Fig_omen}(c) and \ref{Fig_omen}(d), one can find that by harnessing the unique features of the squeezed COM systems (with, e.g., squeezing strength $r_{d}=\textrm{0.25}$), a considerable enhancement of both bipartite light-mirror entanglement and full tripartite entanglement can be achieved by about $\textrm{2}$ or $\textrm{5}$ times, respectively. Moreover, as shown in Figs.\,\ref{Fig_omen}(e) and \ref{Fig_omen}(f), the degree of such bipartite and tripartite entanglement also relies on the squeezing reference angle $\theta_{d}$ and can reach their maximum values at the optimal angle $\theta_{d}=\pi$ or $3\pi$, which is consistent with the previous discussions in Fig.\,\ref{Fig_schematic}. This result implies that for the sake of enhancing entanglement in squeezed COM systems, one should properly choose parameters both for squeezing strength $r_{d}$ and for squeezing reference angle $\theta_{d}$.

Finally, we present how to regulate the behaviors of bipartite and tripartite entanglement involving two mechanical mirrors. For this purpose, we plot the corresponding entanglement measures and the associated reconstructed Wigner function in Fig.\,\ref{Fig_mmen}, where we have altered the following parameters as: $\gamma_{m,j}/\omega_{m}=\textrm{0.2}$, $g_{j}/\omega_{m}=\textrm{0.1}$, $\bar{n}_{m,j}=\textrm{0.05}$. Specifically, as shown in Figs.\,\ref{Fig_mmen}(a)-\ref{Fig_mmen}(c), the logarithmic negativity $E_{\mathcal{N}}^{\,\textit{b}_{\textit{1}}|\textit{b}_{\textit{2}}}$, $E_{\mathcal{N}}^{\,\textit{c}_{\textit{s}}|\textit{b}_{\textit{1}}}$, and the minimum residual contangle $\mathcal{R}^{\textrm{min}}_{\tau}$ are plotted as a function of  the optical detuning $\Delta_{c}$ for different values of $r_{d}$, with $\theta_{d}=\pi$, $\delta r=\textrm{0}$, and $\delta\theta=\pi$. It is seen that in the case of a conventional COM system without any intracavity squeezing (i.e., $r_{d}=\textrm{0}$), the bipartite mirror-mirror entanglement is not present, i.e., $E_{\mathcal{N}}^{\,\textit{b}_{\textit{1}}|\textit{b}_{\textit{2}}}=\textrm{0}$, whereas the bipartite light-mirror entanglement exists and reaches its maximum at $\Delta_{c}/\omega_{m}=\textrm{0.55}$, i.e., $E_{\mathcal{N}}^{\,\textit{c}_{\textit{s}}|\textit{b}_{\textit{1}}}=\textrm{0.059}$. As a result, the full tripartite entanglement is too weak in this case. In contrast, it is remarkable that by using a squeezed COM system with squeezing strength $r_{d}=\textrm{0.6}$ and squeezing reference angle $\theta_{d}=\pi$, the bipartite mirror-mirror entanglement can emerge, where the associated $E_{\mathcal{N}}^{\,\textit{b}_{\textit{1}}|\textit{b}_{\textit{2}}}$ achieves a considerable maximum value of about $E_{\mathcal{N}}^{\,\textit{b}_{\textit{1}}|\textit{b}_{\textit{2}}}=\textrm{0.11}$. Correspondingly, the full tripartite entanglement is significantly enhanced as well, and the profile of $\mathcal{R}^{\textrm{min}}_{\tau}$ is characterized by a sharp peak with its maximum value of about $\mathcal{R}^{\textrm{min}}_{\tau}=\textrm{0.044}$ around the optical detuning $\Delta_{c}/\omega_{m}=\textrm{0.2}$. Moreover, it is also seen that the light-mirror entanglement $E_{\mathcal{N}}^{\,\textit{c}_{\textit{s}}|\textit{b}_{\textit{1}}}$ is suppressed when the mirror-mirror entanglement emerges, resulting from the complementary distribution of entanglement. To clearly see this phenomenon, we further plot the dependence of the entanglement measures on squeezing parameters $r_{d}$ and $\theta_{d}$ in the inserted figure of Figs.\,\ref{Fig_mmen}(a)-\ref{Fig_mmen}(c). The result indicates that in the case of a small squeezing strength, the light field tends to be strongly entangled with the mirror, and there is neither mirror-mirror entanglement nor full tripartite entanglement. However, as the squeezing strength increases, the initial light-mirror entanglement can be partially or even totally transferred to the mirror-mirror subsystem, which leads to a significant enhancement of the full tripartite entanglement. Moreover, it is seen that due to the complementary distribution of the entanglement, the initial light-mirror entanglement $E_{\mathcal{N}}^{\,\textit{c}_{\textit{s}}|\textit{b}_{\textit{1}}}$ is suppressed as it is partially transferred to the mirror-mirror subsystem. We also show the dependence of the entanglement measures on squeezing parameters $r_{d}$ and $\theta_{d}$ in the inserted figure of Figs.\,\ref{Fig_mmen}(a)-\ref{Fig_mmen}(c), which demonstrates a more clear complementary distribution of the entanglement. In Figs.\,\ref{Fig_mmen}(d)-\ref{Fig_mmen}(f), we plot a selection of characteristic projections of the reconstructed Wigner function $\textrm{W}(\psi)$ on two-dimensional subspaces with $\Delta_{c}/\omega_{m}=\textrm{0.55}$, $\delta r=\textrm{0}$, and $\delta\theta=\pi$, showing the quadrature statistics for the two mechanical mirrors with respect to different squeezing parameters. The ellipse (circle) with solid (dashed) line corresponds to a drop by $1/\textrm{e}$ of the maximum value of $\textrm{W}(\psi)$ for the steady (vacuum) state of the system. Note that when the boundary of the ellipse enters the circle, it indicates the existence of a squeezed state, where the variance of the relevant steady state of the system is smaller than that of a vacuum state. In the absence of intracavity squeezing, i.e., $r_{d}=\textrm{0}$, it is found that none of the characteristic projections of $\textrm{W}(\psi)$ is characterized by squeezing correlation for the quadrature pairs of the two mechanical mirrors [see Fig.\,\ref{Fig_mmen}(d)]. Meanwhile, in the presence of intracavity squeezing, i.e., $r_{d}=\textrm{0.6}$, it is found that the emergence of quadrature squeezing depends on the option of squeezing reference angle $\theta_{d}$. For example, when choosing $\theta_{d}=\textrm{0}$, the projections of $\textrm{W}(\psi)$ for all possible permutations of quadrature pairs is consistent with a Gaussian distribution of large variance, resulting from the phase-preserving amplification of the vacuum state [see Fig.\,\ref{Fig_mmen}(e)]. Nevertheless, when choosing $\theta_{d}=\pi$, the projection of $\textrm{W}(\psi)$ for local quadrature pairs $\{Q_{1},P_{1}\}$, $\{Q_{2},P_{2}\}$ and for cross-quadrature pairs $\{P_{1},P_{2}\}$ are squeezed along the diagonal direction, indicating that both single-mode and two-mode squeezing correlations emerge for the associated mechanical mirrors [see Fig.\,\ref{Fig_mmen}(f)]. These results are consistent with the features of the mirror-mirror entanglement previously shown in Figs.\,\ref{Fig_mmen}(a)-\ref{Fig_mmen}(c).

\section{Conclusion}\label{secIV}

In summary, we have presented a theoretical proposal to achieve enhanced and controllable light-mirror, mirror-mirror and full tripartite entanglement in a squeezed COM system, which consists of a FP cavity with two movable mirrors, a nonlinear $\chi^{(2)}$ medium, and an injected broadband squeezed laser field. The nonlinear $\chi^{(2)}$ medium serves as an OPA that enables a squeezed intracavity mode for the COM system. Meanwhile, the injected broadband squeezed laser field can act as a squeezed vacuum reservoir. We show that the interplay of the squeezed intracavity mode and the squeezed vacuum reservoir allows us to effectively enhance the light-mirror interaction while suppress the optical input noises by choosing proper squeezing parameters, where the fulfillment of the phase-matched condition that achieves at $\delta r=0$ and $\delta\theta=\pm n\pi~(n=1,3,5,\ldots)$ plays a key role in this procedure. More interestingly, we find that these features of squeezed COM system provide an appealing method to coherently generate, manipulate and enhance the light-mirror, mirror-mirror and full tripartite entanglement, as well as the associated single-mode or two-mode quadrature squeezing. We remark that our work reveals the potential applications of the squeezed COM system to engineer various quantum COM effects, such as photon blockade~\cite{Lu2015PRL,Tang2022PRL,Wang2023OE,Shen2023PRA,Wang2022PRAPP}, quantum squeezing~\cite{Qin2022PRL,Zhang2021PRA,Huang2023PRA}, quantum phase transition~\cite{Lu2018PRAPP}, and optical or mechanical cat state~\cite{Qin2021PRL,Chen2021PRL,Li2023FR}, to name a few. Also, our work holds the promise to provide a controllable quantum resource for a wide range of entanglement-based quantum technologies, including quantum computing~\cite{Zhang2018NSR}, quantum sensing~\cite{Degen2017RMP,Fleury2015NC,Zhao2020SCMPA} and quantum networking~\cite{XuRMP2020,Kimble2008Nature,Kuzyk2018PRX}. Moreover, although we have considered here a specific case of a squeezed COM system, we envision that our proposal can be extended to a lot of other optical platforms to regulate their light-matter interaction, such as those involving atoms~\cite{Dai2016NP,Karg2020Science,Luo2022PRL}, magnons~\cite{Li2018PRL,Huang2022OL}, and superconducting circuits~\cite{Campagne2018PRL,Kurpiers2018Nature}.

\section{Acknowledgment}

The authors thank Wei Qin, Deng-Gao Lai, and Jian Huang for helpful discussions. H.J. is supported by the National Natural Science Foundation of China (NSFC, Grant No. 11935006), the National Key R\&D Program of China (Grant No. 2024YFE0102400), the Hunan Provincial Major Sci-Tech Program (Grant No. 2023ZJ1010) and the Science and Technology Innovation Program of Hunan Province (Grant No. 2020RC4047). L.-M.K. is supported by the NSFC (Grants No. 12247105, 11935006 and 12175060), Hunan Provincial Major Sci-Tech Program (Grant No. 2023ZJ1010), and
Henan Science and Technology Major Project (No. 241100210400). J.-Q.L. was supported in part by the NSFC (Grants No.~12175061, No.~12247105, and No.~11935006), the National Key R\&D Program of China (Grant No. 2024YFE0102400), and the Hunan Provincial Major Sci-Tech Program (Grant No.~2023ZJ1010). Y.-F.J. is supported by the NSFC (Grant No. 12147156). Y.W. is supported by the NSFC (Grant No. 12205256) and the Henan Provincial Science and Technology Research Project (Grant No. 232102221001). W.L. is supported by the NSFC (Grant No. 12205092) and the Hunan Provincial Natural Science Foundation of China (Grant No. 2023JJ40208).

\clearpage

\onecolumngrid

\setcounter{equation}{0} \setcounter{figure}{0}
\setcounter{table}{0} \setcounter{page}{1}\setcounter{secnumdepth}{3} \makeatletter
\renewcommand{\theequation}{S\arabic{section}.\arabic{equation}}
\renewcommand{\thefigure}{S\arabic{figure}}
\renewcommand{\bibnumfmt}[1]{[S#1]}
\renewcommand{\citenumfont}[1]{S#1}
\renewcommand\thesection{S\arabic{section}}
\makeatletter
\newcommand{\rmnum}[1]{\romannumeral #1}
\newcommand{\Rmnum}[1]{\expandafter\@slowromancap\romannumeral #1@}
\makeatother

\begin{center}
{\ \ }
\end{center}

\begin{center}
{\large \bf Supporting Information for ``Tripartite quantum entanglement with squeezed optomechanics''}
\end{center}

\vspace{1mm}

\begin{center}
Ya-Feng Jiao,$^{1,2}$ Yun-Lan Zuo,$^{3}$ Yan Wang,$^{1,2}$ Wangjun Lu,$^{4}$ Jie-Qiao Liao, $^{3}$ Le-Man Kuang,$^{2,3,\,*}$ and Hui Jing$^{2,3,\dagger}$
\end{center}

\begin{minipage}[]{16cm}
\small{\it
\centering $^{1}$School of Electronics and Information, Zhengzhou University of Light Industry, Zhengzhou 450001, China \\
\centering $^{2}$Academy for Quantum Science and Technology, Zhengzhou University of Light Industry, Zhengzhou 450002, China \\
\centering $^{3}$Key Laboratory of Low-Dimensional Quantum Structures and Quantum Control of Ministry of Education,  \\
\centering Department of Physics and Synergetic Innovation Center for Quantum Effects and Applications, \\
\centering Hunan Normal University, Changsha 410081, China \\
\centering $^{4}$Institute of Engineering Education and Engineering Culture Innovation and Department of Maths and Physics, \\
\centering Hunan Institute of Engineering, Xiangtan 411104, China \\
\centering $*$ lmkuang@hunnu.edu.cn \\
\centering $\dagger$ jinghui73@foxmail.com \\
}
\end{minipage}
\setcounter{section}{0}

\section{Derivation of the effective Hamiltonian and master equation} \label{AppendixA}
The Hamiltonian of the cavity mode $H_{\textrm{c}}$ in Eq.\,(1) includes two quadratic terms describing the nonlinear optical interaction induced by OPA. To introduce the squeezed cavity mode for $H_{\textrm{c}}$, one can perform a unitary Bogoliubov transformation with $S(\eta_{d})=\exp[(-\eta_{d}\hat{c}^{\dagger2}+\eta_{d}^{\ast}\hat{c}^{2})/2]$, where $\eta_{d}=r_{d}\exp{(-i\theta_{d})}$ is a complex number with a squeezing strength $r_{d}$ and a squeezing reference angle $\theta_{d}$. By employing the Bogoliubov transformation, the system Hamiltonian in Eq.\,(1) becomes
\begin{align}
\tilde{H}=&S^{\dagger}(\eta_{d})HS(\eta_{d}) \notag \\
=&\omega_{m,1}\hat{b}_{1}^{\dagger}\hat{b}_{1}+\omega_{m,2}\hat{b}_{2}^{\dagger}\hat{b}_{2}+\chi(e^{-i\varphi}\hat{b}_{1}^{\dagger}\hat{b}_{2}+e^{i\varphi}\hat{b}_{2}^{\dagger}\hat{b}_{1})+F_{1}(\hat{b}_{1}^{\dagger}+\hat{b}_{1})+F_{2}(\hat{b}_{2}^{\dagger}+\hat{b}_{2})
\notag \\
&+\left[\Delta_{c}\cosh(2r_{d})-2\Xi_{d}\sinh(2r_{d})\right]\hat{c}_{s}^{\dagger}\hat{c}_{s}
\notag \\
&+\left[\Xi_{d}\cosh(2r_{d})-\dfrac{\Delta_{c}}{2}\sinh(2r_{d})\right]e^{-i\theta_{d}}\hat{c}_{s}^{\dagger2}
\notag \\
&+\left[\Xi_{d}\cosh(2r_{d})-\dfrac{\Delta_{c}}{2}\sinh(2r_{d})\right]e^{i\theta_{d}}\hat{c}_{s}^{2}
\notag \\
&-g_{1}\left[\cosh(2r_{d})\hat{c}_{s}^{\dagger}\hat{c}_{s}-\dfrac{1}{2}\sinh(2r_{d})(e^{-i\theta_{d}}\hat{c}_{s}^{\dagger2}+e^{i\theta_{d}}\hat{c}_{s}^{2})+\sinh^{2}(r_{d})\right](\hat{b}_{1}^{\dagger}+\hat{b}_{1})
\notag \\
&-g_{2}\left[\cosh(2r_{d})\hat{c}_{s}^{\dagger}\hat{c}_{s}-\dfrac{1}{2}\sinh(2r_{d})(e^{-i\theta_{d}}\hat{c}_{s}^{\dagger2}+e^{i\theta_{d}}\hat{c}_{s}^{2})+\sinh^{2}(r_{d})\right](\hat{b}_{2}^{\dagger}+\hat{b}_{2})
\notag \\
&+i\varepsilon_{d}\cosh(r_{d})(\hat{c}_{s}^{\dagger}-\hat{c}_{s})
+i\varepsilon_{d}\sinh(r_{d})(e^{-i\theta_{d}}\hat{c}_{s}^{\dagger}-e^{i\theta_{d}}\hat{c}_{s})
\notag \\
&+\Delta_{c}\sinh^{2}(r_{d})-\Xi_{d}\sinh(2r_{d}).
\end{align}
To diagonalize the Hamiltonian of the cavity mode and cancel the force induced by the parametric amplification, we set
\begin{align}
&\Xi_{d}\cosh(2r_{d})-\dfrac{\Delta_{c}}{2}\sinh(2r_{d})=0,
\notag \\
&F_{j}-g_{j}\sinh^{2}(r_{d})=0,
\end{align}
and then one can obtain $r_{d}=(1/4)\ln[(\Delta_{c}+2\Xi_{d})/(\Delta_{c}-2\Xi_{d})]$ and
$F_{j}=g_{j}\sinh^{2}(r_{d})$. As a result, the effective Hamiltonian of the system can be obtained as
\begin{align}
\tilde{H}\!=
&\omega_{s}\hat{c}_{s}^{\dagger}\hat{c}_{s}\!+\!\omega_{m,1}\hat{b}_{1}^{\dagger}\hat{b}_{1}\!+\!\omega_{m,2}\hat{b}_{2}^{\dagger}\hat{b}_{2}
\!+\!\chi(e^{-i\varphi}\hat{b}_{1}^{\dagger}\hat{b}_{2}\!+\!e^{i\varphi}\hat{b}_{2}^{\dagger}\hat{b}_{1})
\notag \\
&-\!\zeta_{s,1}\hat{c}_{s}^{\dagger}\hat{c}_{s}(\hat{b}_{1}^{\dagger}\!+\!\hat{b}_{1})\!+\!\dfrac{\zeta_{p,1}}{2}(e^{-i\theta_{d}}\hat{c}_{s}^{\dagger2}\!+\!e^{i\theta_{d}}\hat{c}_{s}^{2})(\hat{b}_{1}^{\dagger}\!+\!\hat{b}_{1})
\notag \\
&-\!\zeta_{s,2}\hat{c}_{s}^{\dagger}\hat{c}_{s}(\hat{b}_{2}^{\dagger}\!+\!\hat{b}_{2})\!+\!\dfrac{\zeta_{p,2}}{2}(e^{-i\theta_{d}}\hat{c}_{s}^{\dagger2}\!+\!e^{i\theta_{d}}\hat{c}_{s}^{2})(\hat{b}_{2}^{\dagger}\!+\!\hat{b}_{2})
\notag \\
&+\!i\varepsilon_{d}\!\cosh(r_{d})(\hat{c}_{s}^{\dagger}\!-\!\hat{c}_{s})
\!+\!i\varepsilon_{d}\!\sinh(r_{d})(e^{-i\theta_{d}}\hat{c}_{s}^{\dagger}\!-\!e^{i\theta_{d}}\hat{c}_{s}),
\end{align}
where
\begin{align}
\nonumber
&\omega_{s}=\Delta_{c}\cosh(2r_{d})-2\Xi_{d}\sinh(2r_{d})=(\Delta_{c}-2\Xi_{d})\exp(2r_{d}),\\ \nonumber
&\zeta_{s,j}=g_{j}\cosh(2r_{d})=\dfrac{g_{j}\Delta_{c}}{\sqrt{\Delta_{c}^{2}-4\Xi_{d}^{2}}},\notag\\
&\zeta_{p,j}=g_{j}\sinh(2r_{d})=\dfrac{2g_{j}G}{\sqrt{\Delta_{c}^{2}-4\Xi_{d}^{2}}},\notag\\
&C=\Delta_{c}\sinh^{2}(r_{d})-G\sinh(2r_{d}).
\end{align}
Correspondingly, the Lindblad operators under the Bogoliubov transformation becomes
\begin{align}
&S^{\dagger}(\eta_{d})\mathcal{D}[\hat{c}]\rho S(\eta_{d}) \notag \\
=&2\cosh^{2}(r_{d})\hat{c}_{s}\tilde{\rho}\hat{c}_{s}^{\dagger}
+2\sinh^{2}(r_{d})\hat{c}_{s}^{\dagger}\tilde{\rho}\hat{c}_{s}
-2\sinh(r_{d})\cosh(r_{d})(e^{-i\theta_{d}}\hat{c}_{s}^{\dagger}\tilde{\rho}\hat{c}_{s}^{\dagger}+e^{i\theta_{d}}\hat{c}_{s}\tilde{\rho}\hat{c}_{s})
\notag \\
&-\left[\cosh^{2}(r_{d})\hat{c}_{s}^{\dagger}\hat{c}_{s}\tilde{\rho}
+\sinh^{2}(r_{d})\hat{c}_{s}\hat{c}_{s}^{\dagger}\tilde{\rho}
-\sinh(r_{d})\cosh(r_{d})(e^{-i\theta_{d}}\hat{c}_{s}^{\dagger}\hat{c}_{s}^{\dagger}\tilde{\rho}+e^{i\theta_{d}}\hat{c}_{s}\hat{c}_{s}\tilde{\rho})\right]
\notag \\
&-\left[\cosh^{2}(r_{d})\tilde{\rho}\hat{c}_{s}^{\dagger}\hat{c}_{s}+\sinh^{2}(r_{d})\tilde{\rho}\hat{c}_{s}\hat{c}_{s}^{\dagger}-\sinh(r_{d})\cosh(r_{d})(e^{-i\theta_{d}}\tilde{\rho}\hat{c}_{s}^{\dagger}\hat{c}_{s}^{\dagger}+e^{i\theta_{d}}\tilde{\rho}\hat{c}_{s}\hat{c}_{s})\right],
\notag \\
\notag \\
&S^{\dagger}\mathcal{G}[\hat{c}]\rho S(\eta_{d}) \notag \\
=&2e^{-2i\theta_{d}}\sinh^{2}(r_{d})\hat{c}_{s}^{\dagger}\tilde{\rho}\hat{c}_{s}^{\dagger}
+2\cosh^{2}(r_{d})\hat{c}_{s}\tilde{\rho}\hat{c}_{s}
-2e^{-i\theta_{d}}\sinh(r_{d})\cosh(r_{d})(\hat{c}_{s}^{\dagger}\tilde{\rho}\hat{c}_{s}+\hat{c}_{s}\tilde{\rho}\hat{c}_{s}^{\dagger})
\notag \\
&-\left[e^{-2i\theta_{d}}\sinh^{2}(r_{d})\hat{c}_{s}^{\dagger}\hat{c}_{s}^{\dagger}\tilde{\rho}+\cosh^{2}(r_{d})\hat{c}_{s}\hat{c}_{s}\tilde{\rho}-e^{-i\theta_{d}}\sinh(r_{d})\cosh(r_{d})(\hat{c}_{s}^{\dagger}\hat{c}_{s}\tilde{\rho}+\hat{c}_{s}\hat{c}_{s}^{\dagger}\tilde{\rho})\right]
\notag \\
&-\left[e^{-2i\theta_{d}}\sinh^{2}(r_{d})\tilde{\rho}\hat{c}_{s}^{\dagger}\hat{c}_{s}^{\dagger}+\cosh^{2}(r_{d})\tilde{\rho}\hat{c}_{s}\hat{c}_{s}-e^{-i\theta_{d}}\sinh(r_{d})\cosh(r_{d})(\tilde{\rho}\hat{c}_{s}^{\dagger}\hat{c}_{s}+\tilde{\rho}\hat{c}_{s}\hat{c}_{s}^{\dagger})\right],
\end{align}
yielding the effective master equation as
\begin{align}
\dot{\tilde{\rho}}=&S^{\dagger}(\eta_{d})\dot{\rho}S(\eta_{d})
\notag \\
=&i\big[\tilde{\rho},\tilde{H}\big]+\frac{\kappa}{2}(N_{s}+1)\mathcal{D}[\hat{c}_{s}]\tilde{\rho}
+\frac{\kappa}{2}N_{s}\mathcal{D}[\hat{c}_{s}^{\dagger}]\tilde{\rho}
-\frac{\kappa}{2}M_{s}\mathcal{G}[\hat{c}_{s}]\tilde{\rho}
-\frac{\kappa}{2}M_{s}^{\ast}\mathcal{G}[\hat{c}_{s}^{\dagger}]\tilde{\rho}
\notag \\
&+\sum_{j=1,2}\dfrac{\gamma_{m,j}}{2}(\bar{n}_{m,j}+1)\mathcal{D}[\hat{b}_{j}]\tilde{\rho}+\dfrac{\gamma_{m,j}}{2}\bar{n}_{m,j}\mathcal{D}[\hat{b}_{j}^{\dagger}]\tilde{\rho},
\end{align}
where
\begin{align}
\nonumber
N_{s}=&\sinh^{2}(r_{d})\cosh^{2}(r_{e})+\cosh^{2}(r_{d})\sinh^{2}(r_{e})
+\dfrac{1}{2}\cos(\theta_{e}-\theta_{d})\sinh(2r_{d})\sinh(2r_{e}),\\ \nonumber
M_{s}=&e^{i\theta_{d}}[\cosh(r_{d})\cosh(r_{e})+e^{-i(\theta_{e}-\theta_{d})}\sinh(r_{d})\sinh(r_{e})]\times[\sinh(r_{d})\cosh(r_{e})+e^{i(\theta_{e}-\theta_{d})}\cosh(r_{d})\sinh(r_{e})].
\end{align}
Here $ N_{s} $ and $ M_{s} $ denote the effective thermal noise and two-photon-correlation strength, respectively, which can be simplified in case of $ r_{d}=r_{e}=r $, i.e.,
\begin{align}
\nonumber
N_{s}=&\dfrac{1}{2}\sinh^{2}(2r)[1+\cos(\theta_{e}-\theta_{d})],\\
M_{s}=&\dfrac{1}{2}e^{i\theta_{d}}\sinh(2r)[1+e^{i(\theta_{e}-\theta_{d})}][\cosh^{2}(r)+e^{-i(\theta_{e}-\theta_{d})}\sinh^{2}(r)].
\end{align}

\section{Derivation of the dynamics of quantum correlation function} \label{AppendixB}
Equations (19) and (20) indicate that the matrix elements satisfy the relation of $V_{kl}=V_{lk}$, and that they are related to the quantum correlation functions for the associated optical and mechanical modes in terms of bosonic operators, i.e.,
\begin{align}
&V_{11}=\text{Re}\left[x_{7}\right]+x_{1}+\frac{1}{2},~
V_{22}=-\text{Re}\left[x_{7}\right]+x_{1}+\frac{1}{2},~
V_{33}=\text{Re}\left[x_{9}\right]+x_{3}+\frac{1}{2},\notag \\
&V_{44}=-\text{Re}\left[x_{9}\right]+x_{3}+\frac{1}{2},~
V_{55}=\text{Re}\left[x_{11}\right]+x_{5}+\frac{1}{2},~
V_{66}=-\text{Re}\left[x_{11}\right]+x_{5}+\frac{1}{2}, \notag\\
&V_{12}=\text{Im}\left[x_{7}\right],~
V_{34}=\text{Im}\left[x_{9}\right],~
V_{56}=\text{Im}\left[x_{11}\right],\notag \\
&V_{13}=\text{Re}\left[x_{13}\right]+\text{Re}\left[x_{15}\right],~
V_{14}=\text{Im}\left[x_{13}\right]-\text{Im}\left[x_{15}\right],~
V_{15}=\text{Re}\left[x_{17}\right]+\text{Re}\left[x_{19}\right],\notag \\
&V_{16}=\text{Im}\left[x_{17}\right]-\text{Im}\left[x_{19}\right],~
V_{23}=\text{Im}\left[x_{13}\right]+\text{Im}\left[x_{15}\right],~
V_{24}=-\text{Re}\left[x_{13}\right]+\text{Re}\left[x_{15}\right],\notag \\
&V_{25}=\text{Im}\left[x_{17}\right]+\text{Im}\left[x_{19}\right],~
V_{26}=-\text{Re}\left[x_{17}\right]+\text{Re}\left[x_{19}\right],~
V_{35}=\text{Re}\left[x_{21}\right]+\text{Re}\left[x_{23}\right],\notag \\
&V_{36}=\text{Im}\left[x_{21}\right]-\text{Im}\left[x_{23}\right],~
V_{45}=\text{Im}\left[x_{21}\right]+\text{Im}\left[x_{23}\right],~
V_{46}=-\text{Re}\left[x_{21}\right]+\text{Re}\left[x_{23}\right],
\label{Vele}
\end{align}
with
\begin{align}
&x_{1}=\langle\hat{c}_{s}^{\dagger}\hat{c}_{s}\rangle,
~x_{2}=\langle\hat{c}_{s}\hat{c}_{s}^{\dagger}\rangle,
~x_{3}=\langle\hat{b}_{1}^{\dagger}\hat{b}_{1}\rangle,
~x_{4}=\langle\hat{b}_{1}\hat{b}_{1}^{\dagger}\rangle,
~x_{5}=\langle\hat{b}_{2}^{\dagger}\hat{b}_{2}\rangle,
~x_{6}=\langle\hat{b}_{2}\hat{b}_{2}^{\dagger}\rangle,
\notag \\
&x_{7}=\langle\hat{c}_{s}\hat{c}_{s}\rangle,
~x_{8}=\langle\hat{c}_{s}^{\dagger}\hat{c}_{s}^{\dagger}\rangle,
~x_{9}=\langle\hat{b}_{1}\hat{b}_{1}\rangle,
~x_{10}=\langle\hat{b}_{1}^{\dagger}\hat{b}_{1}^{\dagger}\rangle,
~x_{11}=\langle\hat{b}_{2}\hat{b}_{2}\rangle,
~x_{12}=\langle\hat{b}_{2}^{\dagger}\hat{b}_{2}^{\dagger}\rangle,
\notag \\
&x_{13}=\langle\hat{c}_{s}\hat{b}_{1}\rangle,
~x_{14}=\langle\hat{c}_{s}^{\dagger}\hat{b}_{1}^{\dagger}\rangle,
~x_{15}=\langle\hat{c}_{s}\hat{b}_{1}^{\dagger}\rangle,
~x_{16}=\langle\hat{c}_{s}^{\dagger}\hat{b}_{1}\rangle,
~x_{17}=\langle\hat{c}_{s}\hat{b}_{2}\rangle,
~x_{18}=\langle\hat{c}_{s}^{\dagger}\hat{b}_{2}^{\dagger}\rangle,
\notag \\
&x_{19}=\langle\hat{c}_{s}\hat{b}_{2}^{\dagger}\rangle,
~x_{20}=\langle\hat{c}_{s}^{\dagger}\hat{b}_{2}\rangle,
~x_{21}=\langle\hat{b}_{1}\hat{b}_{2}\rangle,
~x_{22}=\langle\hat{b}_{1}^{\dagger}\hat{b}_{2}^{\dagger}\rangle,
~x_{23}=\langle\hat{b}_{1}\hat{b}_{2}^{\dagger}\rangle,
~x_{24}=\langle\hat{b}_{1}^{\dagger}\hat{b}_{2}\rangle.
\label{Xele}
\end{align}
By employing the linearized effective master equation (14), the equation of motion of the quantum correlation function in the vector $X$ can be derived as
\begin{align}
\dfrac{d}{dt}\left\langle\hat{c}_{s}^{\dagger}\hat{c}_{s}\right\rangle
=&-\kappa\left\langle\hat{c}_{s}^{\dagger}\hat{c}_{s}\right\rangle
+i\Lambda_{1}\left\langle\hat{c}_{s}^{\dagger}\hat{b}_{1}^{\dagger}\right\rangle
+i\Lambda_{1}\left\langle\hat{c}_{s}^{\dagger}\hat{b}_{1}\right\rangle
-i\Lambda_{1}^{\ast}\left\langle\hat{c}_{s}\hat{b}_{1}^{\dagger}\right\rangle
-i\Lambda_{1}^{\ast}\left\langle\hat{c}_{s}\hat{b}_{1}\right\rangle
+i\Lambda_{2}\left\langle\hat{c}_{s}^{\dagger}\hat{b}_{2}^{\dagger}\right\rangle
\notag \\
&+i\Lambda_{2}\left\langle\hat{c}_{s}^{\dagger}\hat{b}_{2}\right\rangle
-i\Lambda_{2}^{\ast}\left\langle\hat{c}_{s}\hat{b}_{2}^{\dagger}\right\rangle
-i\Lambda_{2}^{\ast}\left\langle\hat{c}_{s}\hat{b}_{2}\right\rangle
+\kappa N_{s},
\end{align}
\begin{align}
\dfrac{d}{dt}\left\langle\hat{c}_{s}\hat{c}_{s}^{\dagger}\right\rangle
=&-\kappa\left\langle\hat{c}_{s}\hat{c}_{s}^{\dagger}\right\rangle
-i\Lambda_{1}^{\ast}\left\langle\hat{c}_{s}\hat{b}_{1}\right\rangle
-i\Lambda_{1}^{\ast}\left\langle\hat{c}_{s}\hat{b}_{1}^{\dagger}\right\rangle
+i\Lambda_{1}\left\langle\hat{c}_{s}^{\dagger}\hat{b}_{1}\right\rangle
+i\Lambda_{1}\left\langle\hat{c}_{s}^{\dagger}\hat{b}_{1}^{\dagger}\right\rangle
-i\Lambda_{2}^{\ast}\left\langle\hat{c}_{s}\hat{b}_{2}\right\rangle
\notag \\
&-i\Lambda_{2}^{\ast}\left\langle\hat{c}_{s}\hat{b}_{2}^{\dagger}\right\rangle
+i\Lambda_{2}\left\langle\hat{c}_{s}^{\dagger}\hat{b}_{2}\right\rangle
+i\Lambda_{2}\left\langle\hat{c}_{s}^{\dagger}\hat{b}_{2}^{\dagger}\right\rangle
+\kappa(N_{s}+1),
\end{align}
\begin{align}
\dfrac{d}{dt}\left\langle\hat{b}_{1}^{\dagger}\hat{b}_{1}\right\rangle
=&-\gamma_{m,1}\left\langle\hat{b}_{1}^{\dagger}\hat{b}_{1}\right\rangle
+i\Lambda_{1}\left\langle\hat{c}_{s}^{\dagger}\hat{b}_{1}^{\dagger}\right\rangle
+i\Lambda_{1}^{\ast}\left\langle\hat{c}_{s}\hat{b}_{1}^{\dagger}\right\rangle
-i\Lambda_{1}\left\langle\hat{c}_{s}^{\dagger}\hat{b}_{1}\right\rangle
-i\Lambda_{1}^{\ast}\left\langle\hat{c}_{s}\hat{b}_{1}\right\rangle
-i\chi e^{-i\varphi}\left\langle\hat{b}_{1}^{\dagger}\hat{b}_{2}\right\rangle
\notag \\
&+i\chi e^{i\varphi}\left\langle\hat{b}_{2}^{\dagger}\hat{b}_{1}\right\rangle
+\gamma_{m,1}\bar{n}_{m,1},
\end{align}
\begin{align}
\dfrac{d}{dt}\left\langle\hat{b}_{1}\hat{b}_{1}^{\dagger}\right\rangle
=&-\gamma_{m,1}\left\langle\hat{b}_{1}\hat{b}_{1}^{\dagger}\right\rangle
-i\Lambda_{1}^{\ast}\left\langle\hat{c}_{s}\hat{b}_{1}\right\rangle
-i\Lambda_{1}\left\langle\hat{c}_{s}^{\dagger}\hat{b}_{1}\right\rangle
+i\Lambda_{1}^{\ast}\left\langle\hat{c}_{s}\hat{b}_{1}^{\dagger}\right\rangle
+i\Lambda_{1}\left\langle\hat{c}_{s}^{\dagger}\hat{b}_{1}^{\dagger}\right\rangle
+i\chi e^{i\varphi}\left\langle\hat{b}_{1}\hat{b}_{2}^{\dagger}\right\rangle
\notag \\
&-i\chi e^{-i\varphi}\left\langle\hat{b}_{2}\hat{b}_{1}^{\dagger}\right\rangle
+\gamma_{m,1}(\bar{n}_{m,1}+1),
\end{align}
\begin{align}
\dfrac{d}{dt}\left\langle\hat{b}_{2}^{\dagger}\hat{b}_{2}\right\rangle
=&-\gamma_{m,2}\left\langle\hat{b}_{2}^{\dagger}\hat{b}_{2}\right\rangle
+i\Lambda_{2}\left\langle\hat{c}_{s}^{\dagger}\hat{b}_{2}^{\dagger}\right\rangle
+i\Lambda_{2}^{\ast}\left\langle\hat{c}_{s}\hat{b}_{2}^{\dagger}\right\rangle
-i\Lambda_{2}\left\langle\hat{c}_{s}^{\dagger}\hat{b}_{2}\right\rangle
-i\Lambda_{2}^{\ast}\left\langle\hat{c}_{s}\hat{b}_{2}\right\rangle
+i\chi e^{-i\varphi}\left\langle\hat{b}_{1}^{\dagger}\hat{b}_{2}\right\rangle
\notag \\
&-i\chi e^{i\varphi}\left\langle\hat{b}_{2}^{\dagger}\hat{b}_{1}\right\rangle
+\gamma_{m,2}\bar{n}_{m,2},
\end{align}
\begin{align}
\dfrac{d}{dt}\left\langle\hat{b}_{2}\hat{b}_{2}^{\dagger}\right\rangle
=&-\gamma_{m,2}\left\langle\hat{b}_{2}\hat{b}_{2}^{\dagger}\right\rangle
-i\lambda_{2}^{\ast}\left\langle\hat{c}_{s}\hat{b}_{2}\right\rangle
-i\lambda_{2}\left\langle\hat{c}_{s}^{\dagger}\hat{b}_{2}\right\rangle
+i\lambda_{2}^{\ast}\left\langle\hat{c}_{s}\hat{b}_{2}^{\dagger}\right\rangle
+i\lambda_{2}\left\langle\hat{c}_{s}^{\dagger}\hat{b}_{2}^{\dagger}\right\rangle
-i\chi e^{i\varphi}\left\langle\hat{b}_{1}\hat{b}_{2}^{\dagger}\right\rangle
\notag \\
&+i\chi e^{-i\varphi}\left\langle\hat{b}_{2}\hat{b}_{1}^{\dagger}\right\rangle
+\gamma_{m,2}(\bar{n}_{m,2}+1),
\end{align}
\begin{align}
\dfrac{d}{dt}\left\langle\hat{c}_{s}\hat{c}_{s}\right\rangle
=-(2i\Delta_{s}+\kappa)\left\langle\hat{c}_{s}\hat{c}_{s}\right\rangle
+2i\Lambda_{1}\left\langle\hat{c}_{s}\hat{b}_{1}^{\dagger}\right\rangle
+2i\Lambda_{1}\left\langle\hat{c}_{s}\hat{b}_{1}\right\rangle
+2i\Lambda_{2}\left\langle\hat{c}_{s}\hat{b}_{2}^{\dagger}\right\rangle
+2i\Lambda_{2}\left\langle\hat{c}_{s}\hat{b}_{2}\right\rangle
+\kappa M_{s}^{\ast},
\end{align}
\begin{align}
\dfrac{d}{dt}\left\langle\hat{c}_{s}^{\dagger}\hat{c}_{s}^{\dagger}\right\rangle
=(2i\Delta_{s}-\kappa)\left\langle\hat{c}_{s}^{\dagger}\hat{c}_{s}^{\dagger}\right\rangle
-2i\Lambda_{1}^{\ast}\left\langle\hat{c}_{s}^{\dagger}\hat{b}_{1}\right\rangle
-2i\Lambda_{1}^{\ast}\left\langle\hat{c}_{s}^{\dagger}\hat{b}_{1}^{\dagger}\right\rangle
-2i\Lambda_{2}^{\ast}\left\langle\hat{c}_{s}^{\dagger}\hat{b}_{2}\right\rangle
-2i\Lambda_{2}^{\ast}\left\langle\hat{c}_{s}^{\dagger}\hat{b}_{2}^{\dagger}\right\rangle
+\kappa M_{s},
\end{align}
\begin{align}
\dfrac{d}{dt}\left\langle\hat{b}_{1}\hat{b}_{1}\right\rangle
=-(2i\omega_{m,1}+\gamma_{m,1})\left\langle\hat{b}_{1}\hat{b}_{1}\right\rangle
+2i\Lambda_{1}\left\langle\hat{c}_{s}^{\dagger}\hat{b}_{1}\right\rangle
+2i\Lambda_{1}^{\ast}\left\langle\hat{c}_{s}\hat{b}_{1}\right\rangle
-2i\chi e^{-i\varphi}\left\langle\hat{b}_{1}\hat{b}_{2}\right\rangle,
\end{align}
\begin{align}
\dfrac{d}{dt}\left\langle\hat{b}_{1}^{\dagger}\hat{b}_{1}^{\dagger}\right\rangle
=(2i\omega_{m,1}-\gamma_{m,1})\left\langle\hat{b}_{1}^{\dagger}\hat{b}_{1}^{\dagger}\right\rangle
-2i\Lambda_{1}^{\ast}\left\langle\hat{c}_{s}\hat{b}_{1}^{\dagger}\right\rangle
-2i\Lambda_{1}\left\langle\hat{c}_{s}^{\dagger}\hat{b}_{1}^{\dagger}\right\rangle
+2i\chi e^{i\varphi}\left\langle\hat{b}_{1}^{\dagger}\hat{b}_{2}^{\dagger}\right\rangle,
\end{align}
\begin{align}
\dfrac{d}{dt}\left\langle\hat{b}_{2}\hat{b}_{2}\right\rangle
=-(2i\omega_{m,2}+\gamma_{m,2})\left\langle\hat{b}_{2}\hat{b}_{2}\right\rangle
+2i\Lambda_{2}\left\langle\hat{c}_{s}^{\dagger}\hat{b}_{2}\right\rangle
+2i\Lambda_{2}^{\ast}\left\langle\hat{c}_{s}\hat{b}_{2}\right\rangle
-2i\chi e^{i\varphi}\left\langle\hat{b}_{1}\hat{b}_{2}\right\rangle,
\end{align}
\begin{align}
\dfrac{d}{dt}\left\langle\hat{b}_{2}^{\dagger}\hat{b}_{2}^{\dagger}\right\rangle
=(2i\omega_{m,2}-\gamma_{m,2})\left\langle\hat{b}_{2}^{\dagger}\hat{b}_{2}^{\dagger}\right\rangle
-2i\Lambda_{2}^{\ast}\left\langle\hat{c}_{s}\hat{b}_{2}^{\dagger}\right\rangle
-2i\Lambda_{2}\left\langle\hat{c}_{s}^{\dagger}\hat{b}_{2}^{\dagger}\right\rangle
+2i\chi e^{-i\varphi}\left\langle\hat{b}_{1}^{\dagger}\hat{b}_{2}^{\dagger}\right\rangle,
\end{align}
\begin{align}
\dfrac{d}{dt}\left\langle\hat{c}_{s}\hat{b}_{1}\right\rangle
=&-[(i\Delta_{s}+\omega_{m,1})+\frac{\kappa+\gamma_{m,1}}{2}]\left\langle\hat{c}_{s}\hat{b}_{1}\right\rangle
+i\Lambda_{1}\left\langle\hat{c}_{s}^{\dagger}\hat{c}_{s}\right\rangle
+i\Lambda_{1}\left\langle\hat{b}_{1}\hat{b}_{1}^{\dagger}\right\rangle
+i\Lambda_{1}\left\langle\hat{b}_{1}\hat{b}_{1}\right\rangle
\notag \\
&+i\Lambda_{1}^{\ast}\left\langle\hat{c}_{s}\hat{c}_{s}\right\rangle
+i\Lambda_{2}\left\langle\hat{b}_{1}\hat{b}_{2}^{\dagger}\right\rangle
+i\Lambda_{2}\left\langle\hat{b}_{1}\hat{b}_{2}\right\rangle
-i\chi e^{-i\varphi}\left\langle\hat{c}_{s}\hat{b}_{2}\right\rangle,
\end{align}
\begin{align}
\dfrac{d}{dt}\left\langle\hat{c}_{s}^{\dagger}\hat{b}_{1}^{\dagger}\right\rangle
=&[(i\Delta_{s}+\omega_{m,1})-\frac{\kappa+\gamma_{m,1}}{2}]\left\langle\hat{c}_{s}^{\dagger}\hat{b}_{1}^{\dagger}\right\rangle
-i\Lambda_{1}^{\ast}\left\langle\hat{c}_{s}\hat{c}_{s}^{\dagger}\right\rangle
-i\Lambda_{1}^{\ast}\left\langle\hat{b}_{1}^{\dagger}\hat{b}_{1}\right\rangle
-i\Lambda_{1}^{\ast}\left\langle\hat{b}_{1}^{\dagger}\hat{b}_{1}^{\dagger}\right\rangle
\notag \\
&-i\Lambda_{1}\left\langle\hat{c}_{s}^{\dagger}\hat{c}_{s}^{\dagger}\right\rangle
-i\Lambda_{2}^{\ast}\left\langle\hat{b}_{1}^{\dagger}\hat{b}_{2}\right\rangle
-i\Lambda_{2}^{\ast}\left\langle\hat{b}_{1}^{\dagger}\hat{b}_{2}^{\dagger}\right\rangle
+i\chi e^{i\varphi}\left\langle\hat{c}_{s}^{\dagger}\hat{b}_{2}^{\dagger}\right\rangle,
\end{align}
\begin{align}
\dfrac{d}{dt}\left\langle\hat{c}_{s}\hat{b}_{1}^{\dagger}\right\rangle
=&-[i(\Delta_{s}-\omega_{m,1})+\frac{\kappa+\gamma_{m,1}}{2}]\left\langle\hat{c}_{s}\hat{b}_{1}^{\dagger}\right\rangle
+i\Lambda_{1}\left\langle\hat{b}_{1}^{\dagger}\hat{b}_{1}^{\dagger}\right\rangle
-i\Lambda_{1}\left\langle\hat{c}_{s}^{\dagger}\hat{c}_{s}\right\rangle
+i\Lambda_{1}\left\langle\hat{b}_{1}^{\dagger}\hat{b}_{1}\right\rangle
\notag \\
&-i\Lambda_{1}^{\ast}\left\langle\hat{c}_{s}\hat{c}_{s}\right\rangle
+i\Lambda_{2}\left\langle\hat{b}_{1}^{\dagger}\hat{b}_{2}^{\dagger}\right\rangle
+i\Lambda_{2}\left\langle\hat{b}_{1}^{\dagger}\hat{b}_{2}\right\rangle
+i\chi e^{i\varphi}\left\langle\hat{c}_{s}\hat{b}_{2}^{\dagger}\right\rangle,
\end{align}
\begin{align}
\dfrac{d}{dt}\left\langle\hat{c}_{s}^{\dagger}\hat{b}_{1}\right\rangle
=&[i(\Delta_{s}-\omega_{m,1})-\frac{\kappa+\gamma_{m,1}}{2}]\left\langle\hat{c}_{s}^{\dagger}\hat{b}_{1}\right\rangle
-i\Lambda_{1}^{\ast}\left\langle\hat{b}_{1}\hat{b}_{1}\right\rangle
+i\Lambda_{1}^{\ast}\left\langle\hat{c}_{s}\hat{c}_{s}^{\dagger}\right\rangle
-i\Lambda_{1}^{\ast}\left\langle\hat{b}_{1}\hat{b}_{1}^{\dagger}\right\rangle
\notag \\
&+i\Lambda_{1}\left\langle\hat{c}_{s}^{\dagger}\hat{c}_{s}^{\dagger}\right\rangle
-i\Lambda_{2}^{\ast}\left\langle\hat{b}_{1}\hat{b}_{2}\right\rangle
-i\Lambda_{2}^{\ast}\left\langle\hat{b}_{1}\hat{b}_{2}^{\dagger}\right\rangle
-i\chi e^{-i\varphi}\left\langle\hat{c}_{s}^{\dagger}\hat{b}_{2}\right\rangle,
\end{align}
\begin{align}
\dfrac{d}{dt}\left\langle\hat{c}_{s}\hat{b}_{2}\right\rangle
=&-[i(\Delta_{s}+\omega_{m,2})+\frac{\kappa+\gamma_{m,2}}{2}]\left\langle\hat{c}_{s}\hat{b}_{2}\right\rangle
+i\Lambda_{2}\left\langle\hat{c}_{s}^{\dagger}\hat{c}_{s}\right\rangle
+i\Lambda_{2}\left\langle\hat{b}_{2}\hat{b}_{2}^{\dagger}\right\rangle
+i\Lambda_{2}\left\langle\hat{b}_{2}\hat{b}_{2}\right\rangle
\notag \\
&+i\Lambda_{2}^{\ast}\left\langle\hat{c}_{s}\hat{c}_{s}\right\rangle
+i\Lambda_{1}\left\langle\hat{b}_{1}^{\dagger}\hat{b}_{2}\right\rangle
+i\Lambda_{1}\left\langle\hat{b}_{1}\hat{b}_{2}\right\rangle
-i\chi e^{i\varphi}\left\langle\hat{c}_{s}\hat{b}_{1}\right\rangle,
\end{align}
\begin{align}
\dfrac{d}{dt}\left\langle\hat{c}_{s}^{\dagger}\hat{b}_{2}^{\dagger}\right\rangle
=&[i(\Delta_{s}+\omega_{m,2})-\frac{\kappa+\gamma_{m,2}}{2}]\left\langle\hat{c}_{s}^{\dagger}\hat{b}_{2}^{\dagger}\right\rangle
-i\Lambda_{2}^{\ast}\left\langle\hat{c}_{s}\hat{c}_{s}^{\dagger}\right\rangle
-i\Lambda_{2}^{\ast}\left\langle\hat{b}_{2}^{\dagger}\hat{b}_{2}\right\rangle
-i\Lambda_{2}^{\ast}\left\langle\hat{b}_{2}^{\dagger}\hat{b}_{2}^{\dagger}\right\rangle
\notag \\
&-i\Lambda_{2}\left\langle\hat{c}_{s}^{\dagger}\hat{c}_{s}^{\dagger}\right\rangle
-i\Lambda_{1}^{\ast}\left\langle\hat{b}_{1}\hat{b}_{2}^{\dagger}\right\rangle
-i\Lambda_{1}^{\ast}\left\langle\hat{b}_{1}^{\dagger}\hat{b}_{2}^{\dagger}\right\rangle
+i\chi e^{-i\varphi}\left\langle\hat{c}_{s}^{\dagger}\hat{b}_{1}^{\dagger}\right\rangle,
\end{align}
\begin{align}
\dfrac{d}{dt}\left\langle\hat{c}_{s}\hat{b}_{2}^{\dagger}\right\rangle
=&-[i(\Delta_{s}-\omega_{m,2})+\frac{\kappa+\gamma_{m,2}}{2}]\left\langle\hat{c}_{s}\hat{b}_{2}^{\dagger}\right\rangle
+i\Lambda_{2}\left\langle\hat{b}_{2}^{\dagger}\hat{b}_{2}^{\dagger}\right\rangle
-i\Lambda_{2}\left\langle\hat{c}_{s}^{\dagger}\hat{c}_{s}\right\rangle
+i\Lambda_{2}\left\langle\hat{b}_{2}^{\dagger}\hat{b}_{2}\right\rangle
\notag \\
&-i\Lambda_{2}^{\ast}\left\langle\hat{c}_{s}\hat{c}_{s}\right\rangle
+i\Lambda_{1}\left\langle\hat{b}_{1}^{\dagger}\hat{b}_{2}^{\dagger}\right\rangle
+i\Lambda_{1}\left\langle\hat{b}_{1}\hat{b}_{2}^{\dagger}\right\rangle
+i\chi e^{-i\varphi}\left\langle\hat{c}_{s}\hat{b}_{1}^{\dagger}\right\rangle,
\end{align}
\begin{align}
\dfrac{d}{dt}\left\langle\hat{c}_{s}^{\dagger}\hat{b}_{2}\right\rangle
=&[i(\Delta_{s}-\omega_{m,2})-\frac{\kappa+\gamma_{m,2}}{2}]\left\langle\hat{c}_{s}^{\dagger}\hat{b}_{2}\right\rangle
-i\Lambda_{2}^{\ast}\left\langle\hat{b}_{2}\hat{b}_{2}\right\rangle
+i\Lambda_{2}^{\ast}\left\langle\hat{c}_{s}\hat{c}_{s}^{\dagger}\right\rangle
-i\Lambda_{2}^{\ast}\left\langle\hat{b}_{2}\hat{b}_{2}^{\dagger}\right\rangle
\notag \\
&+i\Lambda_{2}\left\langle\hat{c}_{s}^{\dagger}\hat{c}_{s}^{\dagger}\right\rangle
-i\Lambda_{1}^{\ast}\left\langle\hat{b}_{1}\hat{b}_{2}\right\rangle
-i\Lambda_{1}^{\ast}\left\langle\hat{b}_{1}^{\dagger}\hat{b}_{2}\right\rangle
-i\chi e^{i\varphi}\left\langle\hat{c}_{s}^{\dagger}\hat{b}_{1}\right\rangle,
\end{align}
\begin{align}
\dfrac{d}{dt}\left\langle\hat{b}_{1}\hat{b}_{2}\right\rangle
=&-[i(\omega_{m,1}+\omega_{m,2})+\frac{\gamma_{m,1}+\gamma_{m,2}}{2}]\left\langle\hat{b}_{1}\hat{b}_{2}\right\rangle
+i\Lambda_{1}\left\langle\hat{c}_{s}^{\dagger}\hat{b}_{2}\right\rangle
+i\Lambda_{1}^{\ast}\left\langle\hat{c}_{s}\hat{b}_{2}\right\rangle
+i\Lambda_{2}\left\langle\hat{c}_{s}^{\dagger}\hat{b}_{1}\right\rangle
\notag \\
&+i\Lambda_{2}^{\ast}\left\langle\hat{c}_{s}\hat{b}_{1}\right\rangle
-i\chi e^{i\varphi}\left\langle\hat{b}_{1}\hat{b}_{1}\right\rangle
-i\chi e^{-i\varphi}\left\langle\hat{b}_{2}\hat{b}_{2}\right\rangle,
\end{align}
\begin{align}
\dfrac{d}{dt}\left\langle\hat{b}_{1}^{\dagger}\hat{b}_{2}^{\dagger}\right\rangle
=&[i(\omega_{m,1}+\omega_{m,2})-\frac{\gamma_{m,1}+\gamma_{m,2}}{2}]\left\langle\hat{b}_{1}^{\dagger}\hat{b}_{2}^{\dagger}\right\rangle
-i\Lambda_{1}^{\ast}\left\langle\hat{c}_{s}\hat{b}_{2}^{\dagger}\right\rangle
-i\Lambda_{1}\left\langle\hat{c}_{s}^{\dagger}\hat{b}_{2}^{\dagger}\right\rangle
-i\Lambda_{2}^{\ast}\left\langle\hat{c}_{s}\hat{b}_{1}^{\dagger}\right\rangle
\notag \\
&-i\Lambda_{2}\left\langle\hat{c}_{s}^{\dagger}\hat{b}_{1}^{\dagger}\right\rangle
+i\chi e^{-i\varphi}\left\langle\hat{b}_{1}^{\dagger}\hat{b}_{1}^{\dagger}\right\rangle
+i\chi e^{i\varphi}\left\langle\hat{b}_{2}^{\dagger}\hat{b}_{2}^{\dagger}\right\rangle,
\end{align}
\begin{align}
\dfrac{d}{dt}\left\langle\hat{b}_{1}\hat{b}_{2}^{\dagger}\right\rangle
=&-[i(\omega_{m,1}-\omega_{m,2})+\frac{\gamma_{m,1}+\gamma_{m,2}}{2}]\left\langle\hat{b}_{1}\hat{b}_{2}^{\dagger}\right\rangle
+i\Lambda_{1}\left\langle\hat{c}_{s}^{\dagger}\hat{b}_{2}^{\dagger}\right\rangle
+i\Lambda_{1}^{\ast}\left\langle\hat{c}_{s}\hat{b}_{2}^{\dagger}\right\rangle
-i\Lambda_{2}\left\langle\hat{c}_{s}^{\dagger}\hat{b}_{1}\right\rangle
\notag \\
&-i\Lambda_{2}^{\ast}\left\langle\hat{c}_{s}\hat{b}_{1}\right\rangle
+i\chi e^{-i\varphi}\left\langle\hat{b}_{1}\hat{b}_{1}^{\dagger}\right\rangle
-i\chi e^{-i\varphi}\left\langle\hat{b}_{2}\hat{b}_{2}^{\dagger}\right\rangle,
\end{align}
\begin{align}
\dfrac{d}{dt}\left\langle\hat{b}_{1}^{\dagger}\hat{b}_{2}\right\rangle
=&[i(\omega_{m,1}-\omega_{m,2})-\frac{\gamma_{m,1}+\gamma_{m,2}}{2}]\left\langle\hat{b}_{1}^{\dagger}\hat{b}_{2}\right\rangle
-i\Lambda_{1}^{\ast}\left\langle\hat{c}_{s}\hat{b}_{2}\right\rangle
-i\Lambda_{1}\left\langle\hat{c}_{s}^{\dagger}\hat{b}_{2}\right\rangle
+i\Lambda_{2}^{\ast}\left\langle\hat{c}_{s}\hat{b}_{1}^{\dagger}\right\rangle
\notag \\
&+i\Lambda_{2}\left\langle\hat{c}_{s}^{\dagger}\hat{b}_{1}^{\dagger}\right\rangle
-i\chi e^{i\varphi}\left\langle\hat{b}_{1}^{\dagger}\hat{b}_{1}\right\rangle
+i\chi e^{i\varphi}\left\langle\hat{b}_{2}^{\dagger}\hat{b}_{2}\right\rangle.
\end{align}
Then, by rewriting the above time evolution equations in a matrix form with the vector $X$, one can obtain
\begin{align}
\dfrac{d}{dt}X=M\cdot X+N,
\end{align}
where
\begin{align}
&M= \notag \\
&\left(\begin{smallmatrix}
-\kappa & 0 & 0 & 0 & 0 & 0 & 0 & 0 & 0 & 0 & 0 & 0 & -i\Lambda_{1}^{\ast} & i\Lambda_{1} & -i\Lambda_{1}^{\ast} & i\Lambda_{1} & -i\Lambda_{2}^{\ast} & i\Lambda_{2} & -i\Lambda_{2}^{\ast} & i\Lambda_{2} \\ 0 & 0 & 0 & 0 \\ \\
0 & -\kappa & 0 & 0 & 0 & 0 & 0 & 0 & 0 & 0 & 0 & 0 & -i\Lambda_{1}^{\ast} & i\Lambda_{1} & -i\Lambda_{1}^{\ast} & i\Lambda_{1} & -i\Lambda_{2}^{\ast} & i\Lambda_{2} & -i\Lambda_{2}^{\ast} & i\Lambda_{2} \\ 0 & 0 & 0 & 0 \\ \\
0 & 0 & -\gamma_{m,1} & 0 & 0 & 0 & 0 & 0 & 0 & 0 & 0 & 0 & -i\Lambda_{1}^{\ast} & i\Lambda_{1} & i\Lambda_{1}^{\ast} & -i\Lambda_{1} & 0 & 0 & 0 & 0 \\ 0 & 0 & i\nu^{\ast} & -i\nu \\ \\
0 & 0 & 0 & -\gamma_{m,1} & 0 & 0 & 0 & 0 & 0 & 0 & 0 & 0 & -i\Lambda_{1}^{\ast} & i\Lambda_{1} & i\Lambda_{1}^{\ast} & -i\Lambda_{1} & 0 & 0 & 0 & 0 \\ 0 & 0 & i\nu^{\ast} & -i\nu \\ \\
0 & 0 & 0 & 0 & -\gamma_{m,2} & 0 & 0 & 0 & 0 & 0 & 0 & 0 & 0 & 0 & 0 & 0 & -i\Lambda_{2}^{\ast} & i\Lambda_{2} & i\Lambda_{2}^{\ast} & -i\Lambda_{2} \\ 0 & 0 & -i\nu^{\ast} & i\nu \\ \\
0 & 0 & 0 & 0 & 0 & -\gamma_{m,2} & 0 & 0 & 0 & 0 & 0 & 0 & 0 & 0 & 0 & 0 & -i\Lambda_{2}^{\ast} & i\Lambda_{2} & i\Lambda_{2}^{\ast} & -i\Lambda_{2} \\ 0 & 0 & -i\nu^{\ast} & i\nu \\ \\
0 & 0 & 0 & 0 & 0 & 0 & -\Omega_{1,+} & 0 & 0 & 0 & 0 & 0 & 2i\Lambda_{1} & 0 & 2i\Lambda_{1} & 0 & 2i\Lambda_{2} & 0 & 2i\Lambda_{2} & 0 \\ 0 & 0 & 0 & 0 \\ \\
0 & 0 & 0 & 0 & 0 & 0 & 0 & \Omega_{1,-} & 0 & 0 & 0 & 0 & 0 & -2i\Lambda_{1}^{\ast} & 0 & -2i\Lambda_{1}^{\ast} & 0 & -2i\Lambda_{2}^{\ast} & 0 & -2i\Lambda_{2}^{\ast} \\ 0 & 0 & 0 & 0 \\ \\
0 & 0 & 0 & 0 & 0 & 0 & 0 & 0 & -\Omega_{2,+} & 0 & 0 & 0 & 2i\Lambda_{1}^{\ast} & 0 & 0 & 2i\Lambda_{1} & 0 & 0 & 0 & 0 \\ -2i\nu & 0 & 0 & 0 \\ \\
0 & 0 & 0 & 0 & 0 & 0 & 0 & 0 & 0 & \Omega_{2,-} & 0 & 0 & 0 & -2i\Lambda_{1} & -2i\Lambda_{1}^{\ast} & 0 & 0 & 0 & 0 & 0 \\ 0 & 2i\nu^{\ast} & 0 & 0 \\ \\
0 & 0 & 0 & 0 & 0 & 0 & 0 & 0 & 0 & 0 & -\Omega_{3,+} & 0 & 0 & 0 & 0 & 0 & 2i\Lambda_{2}^{\ast} & 0 & 0 & 2i\Lambda_{2} \\ -2i\nu^{\ast} & 0 & 0 & 0 \\ \\
0 & 0 & 0 & 0 & 0 & 0 & 0 & 0 & 0 & 0 & 0 & \Omega_{3,-} & 0 & 0 & 0 & 0 & 0 & -2i\Lambda_{2} & -2i\Lambda_{2}^{\ast} & 0 \\ 0 & 2i\nu & 0 & 0 \\ \\
i\Lambda_{1} & 0 & 0 & i\Lambda_{1} & 0 & 0 & i\Lambda_{1}^{\ast} & 0 & i\Lambda_{1} & 0 & 0 & 0 & -\Omega_{4,+} & 0 & 0 & 0 & -i\nu & 0 & 0 & 0 \\ i\Lambda_{2} & 0 & i\Lambda_{2} & 0 \\ \\
0 & -i\Lambda_{1}^{\ast} & -i\Lambda_{1}^{\ast} & 0 & 0 & 0 & 0 & -i\Lambda_{1} & 0 & -i\Lambda_{1}^{\ast} & 0 & 0 & 0 & \Omega_{4,-} & 0 & 0 & 0 & i\nu^{\ast} & 0 & 0 \\ 0 & -i\Lambda_{2}^{\ast} & 0 & -i\Lambda_{2}^{\ast} \\ \\
-i\Lambda_{1} & 0 & i\Lambda_{1} & 0 & 0 & 0 & -i\Lambda_{1}^{\ast} & 0 & 0 & i\Lambda_{1} & 0 & 0 & 0 & 0 & -\Omega_{5,+} & 0 & 0 & 0 & i\nu^{\ast} & 0 \\ 0 & i\Lambda_{2} & 0 & i\Lambda_{2} \\ \\
0 & i\Lambda_{1}^{\ast} & 0 & -i\Lambda_{1}^{\ast} & 0 & 0 & 0 & i\Lambda_{1} & -i\Lambda_{1}^{\ast} & 0 & 0 & 0 & 0 & 0 & 0 & \Omega_{5,-} & 0 & 0 & 0 & -i\nu \\ -i\Lambda_{2}^{\ast} & 0 & -i\Lambda_{2}^{\ast} & 0 \\ \\
i\Lambda_{2} & 0 & 0 & 0 & 0 & i\Lambda_{2} & i\Lambda_{2}^{\ast} & 0 & 0 & 0 & i\Lambda_{2} & 0 & -i\nu^{\ast} & 0 & 0 & 0 & -\Omega_{6,+} & 0 & 0 & 0 \\ i\Lambda_{1} & 0 & 0 & i\Lambda_{1} \\ \\
0 & -i\Lambda_{2}^{\ast} & 0 & 0 & -i\Lambda_{2}^{\ast} & 0 & 0 & -i\Lambda_{2} & 0 & 0 & 0 & -i\Lambda_{2}^{\ast} & 0 & i\nu & 0 & 0 & 0 & \Omega_{6,-} & 0 & 0 \\ 0 & -i\Lambda_{1}^{\ast} & -i\Lambda_{1}^{\ast} & 0 \\ \\
-i\Lambda_{2} & 0 & 0 & 0 & i\Lambda_{2} & 0 & -i\Lambda_{2}^{\ast} & 0 & 0 & 0 & 0 & i\Lambda_{2} & 0 & 0 & i\nu & 0 & 0 & 0 & -\Omega_{7,+} & 0 \\ 0 & i\Lambda_{1} & i\Lambda_{1} & 0 \\ \\
0 & i\Lambda_{2}^{\ast} & 0 & 0 & 0 & -i\Lambda_{2}^{\ast} & 0 & i\Lambda_{2} & 0 & 0 & -i\Lambda_{2}^{\ast} & 0 & 0 & 0 & 0 & -i\nu^{\ast} & 0 & 0 & 0 & \Omega_{7,-} \\ -i\Lambda_{1}^{\ast} & 0 & 0 & -i\Lambda_{1}^{\ast} \\ \\
0 & 0 & 0 & 0 & 0 & 0 & 0 & 0 & -i\nu^{\ast} & 0 & -i\nu & 0 & i\Lambda_{2}^{\ast} & 0 & 0 & i\Lambda_{2} & i\Lambda_{1}^{\ast} & 0 & 0 & i\Lambda_{1} \\ -\Omega_{8,+} & 0 & 0 & 0 \\
0 & 0 & 0 & 0 & 0 & 0 & 0 & 0 & 0 & i\nu & 0 & i\nu^{\ast} & 0 & -i\Lambda_{2} & -i\Lambda_{2}^{\ast} & 0 & 0 & -i\Lambda_{1} & -i\Lambda_{1}^{\ast} & 0 \\ 0 & \Omega_{8,-} & 0 & 0 \\ \\
0 & 0 & 0 & i\nu & 0 & -i\nu & 0 & 0 & 0 & 0 & 0 & 0 & -i\Lambda_{2}^{\ast} & 0 & 0 & -i\Lambda_{2} & 0 & i\Lambda_{1} & i\Lambda_{1}^{\ast} & 0 \\ 0 & 0 & -\Omega_{9,+} & 0 \\ \\
0 & 0 & -i\nu^{\ast} & 0 & i\nu^{\ast} & 0 & 0 & 0 & 0 & 0 & 0 & 0 & 0 & i\Lambda_{2} & i\Lambda_{2}^{\ast} & 0 & -i\Lambda_{1}^{\ast} & 0 & 0 & -i\Lambda_{1} \\ 0 & 0 & 0 & \Omega_{9,-}
\end{smallmatrix}\right),
\end{align}
\begin{align}
N=\left[\kappa N_{s},\kappa(N_{s}+1),\gamma_{m,1}\bar{n}_{m,1},\gamma_{m,1}(\bar{n}_{m,1}+1),
\gamma_{m,2}\bar{n}_{m,2},\gamma_{m,2}(\bar{n}_{m,2}+1),\kappa M_{s}^{\ast},\kappa M_{s},0,0,0,0,0,0,0,0,0,0,0,0,0,0,0,0\right]^{T},
\end{align}
with
\begin{align}
\Omega_{1,\pm}&=2i\Delta_{s}\pm\kappa, ~
\Omega_{2,\pm}=2i\omega_{m,1}\pm\gamma_{m,1}, ~
\Omega_{3,\pm}=2i\omega_{m,2}\pm\gamma_{m,2}, \notag \\
\Omega_{4,\pm}&=i(\Delta_{s}+\omega_{m,1})\pm\frac{\kappa+\gamma_{m,1}}{2}, ~
\Omega_{5,\pm}=i(\Delta_{s}-\omega_{m,1})\pm\frac{\kappa+\gamma_{m,1}}{2}, \notag \\
\Omega_{6,\pm}&=i(\Delta_{s}+\omega_{m,2})\pm\frac{\kappa+\gamma_{m,2}}{2},~
\Omega_{7,\pm}=i(\Delta_{s}-\omega_{m,2})\pm\frac{\kappa+\gamma_{m,2}}{2},
\end{align}
\begin{align}
\Omega_{8,\pm}&=i(\omega_{m,1}+\omega_{m,2})\pm\frac{\gamma_{m,1}+\gamma_{m,2}}{2}, ~
\Omega_{9,\pm}=i(\omega_{m,1}-\omega_{m,2})\pm\frac{\gamma_{m,1}+\gamma_{m,2}}{2}.
\end{align}

\end{document}